\documentclass[12pt]{article}
\usepackage[utf8]{inputenc}
\usepackage[T1]{fontenc}
\usepackage{bm}
\usepackage{caption}
\usepackage{graphicx}
\usepackage{amsmath}
\usepackage{url} 
\usepackage{amssymb}
\usepackage{amsfonts}
\usepackage{hyperref}
\usepackage{float}
\usepackage{amsthm}
\usepackage{xcolor}
\usepackage{algorithm2e}
\usepackage{authblk}
\usepackage{array,longtable,ltxtable}
\usepackage{enumitem}
\usepackage{appendix}
\usepackage{placeins}
\usepackage{subcaption}
\usepackage{lineno}
\usepackage{multirow}
\usepackage{multicol}
\usepackage{soul}
\usepackage{xspace}

\usepackage{listings}
\usepackage{xcolor}
\usepackage{wrapfig}
\usepackage{lineno}

\usepackage{siunitx}

\definecolor{col_e}{HTML}{fb7041}
\definecolor{col_s}{HTML}{bad1f2}
\definecolor{col_d}{HTML}{559ca6}
\definecolor{col_t}{HTML}{96ccc8}

\graphicspath{{./fig/}} 

\newcommand{\dymond}{\textit{Dymond}\xspace}
\newcommand{\taggen}{\textit{TagGen}\xspace}
\newcommand{\stm}{\textit{STM}\xspace}
\newcommand{\etngen}{\textit{ETN-gen}\xspace}

\usepackage[superscript]{cite}
\usepackage{a4wide}
\usepackage{mathpazo}
\makeatletter
\renewcommand{\maketitle}{\bgroup\setlength{\parindent}{0pt}
\begin{flushleft}
{\Large{\textbf{\@title}}}

{\small \@author}
\end{flushleft}\egroup
}
\makeatother
\renewenvironment{abstract}
{{\bfseries ABSTRACT}
\setlength{\leftmargin}{0mm}
\setlength{\rightmargin}{\leftmargin}%
\relax}
{\endlist}
\usepackage{fancyhdr}
\newcolumntype{?}{!{\vrule width 1.2pt}}

\usepackage{lineno}

\newcommand{\changed}[1]{\textcolor{black}{#1 }}

\title{Generating fine-grained surrogate temporal networks}
\date{\vspace{-5ex}}
\author[1,2]{A.~Longa}
\author[1]{G.~Cencetti}
\author[3,4]{S.~Lehmann}
\author[2]{A.~Passerini}
\author[1]{B.~Lepri}

\affil[1]{Fondazione Bruno Kessler, Trento, Italy}
\affil[2]{University of Trento, Trento, Italy}
\affil[3]{Tecnical University of Denmark, Kongens Lyngby, Denmark}
\affil[4]{Copenhagen Center for Social Data Science, Copenhagen, Denmark}

\affil[$\star$]{Corresponding author: \href{mailto:lepri@fbk.eu}{lepri@
fbk.eu}}

\begin{document}

\maketitle

\begin{abstract}

\textbf{
Temporal networks are essential for modeling and understanding systems whose behavior varies in time, from social interactions to biological systems. 
Often, however, real-world data are prohibitively expensive to collect in a large scale or unshareable due to privacy concerns. A promising way to bypass the problem consists in generating arbitrarily large and anonymized synthetic graphs with the properties of real-world networks, namely `surrogate networks'. Until now, the generation of realistic surrogate temporal networks has remained an open problem, due to the difficulty of capturing both the temporal and topological properties of the input network, as well as their correlations, in a scalable model. Here, we propose a novel and simple method for generating surrogate temporal networks. Our method decomposes the input network into star-like structures evolving in time. Then those structures are used as building blocks to generate a surrogate temporal network. Our model vastly outperforms current methods across multiple examples of temporal networks in terms of both topological and dynamical similarity. We further show that beyond generating realistic interaction patterns, our method is able to capture intrinsic temporal periodicity of temporal networks, all with an execution time lower than competing methods by multiple orders of magnitude. The simplicity of our algorithm makes it easily interpretable, extendable and algorithmically scalable.}
\end{abstract}

\section{Introduction}
\changed{In} the past decade, temporal networks have driven breakthroughs in real world systems across biology, communications, social interactions, and mobility.
\changed{One of the main advantages of temporal networks resides in their ability to capture complex dynamics such as, for instance, diffusion and contagion~\cite{holme2012temporal, han2004, eagle2006, chechik2008, balcan2009, cattuto2010, starnini2013topological, corsi2018, lambiotte2019, ciaperoni2020}. 
Here we assume that a temporal network is represented in discrete time with each time step corresponding to a static graph, also referred to as a `layer' of the network.}
In order to model realistic dynamics, it is often necessary to employ large temporal networks, including a large number of nodes and \changed{long time intervals, i.e. many temporal layers}~\cite{starnini2012random,rocha2017sampling,cencetti2021digital}.
Many state-of-the-art temporal datasets, however, are limited both in the number of \changed{nodes} and in the number of temporal layers\cite{cattuto2010, isella2011, stehle2011simulation, aharony2011social, sapiezynski2019interaction}.
When the available data are insufficient -- \changed{e.g.~to simulate long-term effects of epidemics} -- datasets are extended by simply repeating the same temporal sequence multiple times, a procedure which is known to result in biases~\cite{stehle2011simulation}.
An appealing solution to the problem of insufficient data is to use \textit{surrogate temporal networks}~\cite{presigny2021building}.
Surrogate temporal networks are synthetic datasets which mimic the real-world temporal patterns relevant for a desired use-case.
\changed{Real networks are indeed known to be characterized by typical patterns of interactions, different in different domains (social, biological, infrastructural, etc.), which can be often recognized and delineated~\cite{mollgaard2017correlations, kikas2013bursty} and the role of surrogate networks is to try to reproduce them.}
The surrogates can \changed{be designed to} involve the desired number of nodes and number of temporal layers, where the actual dynamics are known through smaller studies or via available small datasets.
Moreover, in the case of privacy sensitive data, such as fine-grained records of social interactions~\cite{cretu2022}, surrogate data can \changed{be generated so as to be freely shareable.}
Over the past years, a large number of successful algorithms for \textit{static} network generation have been proposed~\cite{bois2015probabilistic, coscia2020multiplex}; however, extending these models to the dynamic regime has proven prohibitively difficult, due to greatly increased complexity introduced by the temporal dimension.

Indeed, it has become clear that temporal networks are characterized by a highly non-trivial interplay between the instantaneous network topology  at a given time (adjacency, degree distribution, clustering, etc.) and the temporal activation of nodes and links -- how each connection changes over time (duration of interactions, patterns by which new links appear and old ones disappear, etc.).
From the perspective of an individual node, these two dimensions imply that models must take into account $(i)$ time, i.e. the history of what has occurred in the preceding timesteps and $(ii)$ instantaneous local topology, i.e. the current activation of the neighboring nodes. 
The scientific literature is full of studies focusing on the spatial dimension but unable to take into account possible temporal correlations~\cite{barabasi1999emergence,krapivsky2000connectivity,dorogovtsev2000structure,bianconi2001competition,d2007emergence,papadopoulos2012popularity}, or -- alternatively -- works dedicated to model the behavior of individual nodes in time (for example activity driven models~\cite{perra2012activity, starnini2013topological}) which do not \changed{aim to} reproduce realistic network topologies~\cite{gauvin2018randomized}. 
There exist models for link prediction that try to combine temporal and topological dimensions by using small local temporal patterns~\cite{berlingerio2009mining} or building over a backbone of significant links~\cite{presigny2021building}. However, there is currently a dearth of models for generating surrogate networks from scratch that are able to take into account the two dimensions simultaneously.
The few works, that do this rely on temporal motifs, like \dymond~\cite{zeno2021dymond} and \stm~\cite{purohit2018temporal}, or on deep learning like \taggen~\cite{zhou2020data}. 
These three models described in detail in \textit{Methods}, represent the state-of-the-art. 
\changed{All these models however suffer from some limitations and some of the characteristics of the original networks are not always well reproduced by the surrogate networks.}

\changed{In this work we propose an alternative method that is particularly efficient in reproducing temporal networks characterized by high temporal resolution and we test it with a wide range of topological and dynamical measures, comparing it with the above cited approaches.}
\changed{The method that we propose is based on temporal motifs that are defined with an `egocentric' perspective. Conceptually, we collect the local interactions of each node for a small number of time steps in a real network, the \textit{egocentric temporal neighborhood}, and we consider them as representative of that network of interactions. We then use them as building blocks to generate a new synthetic network.}

A major advantage of the egocentric perspective (that ignores connections among neighbors of an ego node) is that it allows us to linearize the concept of node neighborhood sidestepping the subgraph isomorphism problem~\cite{graphiso2020}, \changed{that often represents a bottleneck for the algorithms based on motifs}. This makes the generation process fast and scalable  both in terms of the number of nodes and the number of temporal snapshots. 
Speed turns out to be a fundamental feature, because the other existing methods rely on algorithms of considerably higher complexity that prevent those methods from scaling to even moderately-sized networks.

We test the method, named
\textit{Egocentric Temporal Neighborhood Generator (ETN-gen)}, on a range of different temporal networks. 
In our testing we mainly use social interactions datasets because of richness and availability of these datasets, but the method is general and can be used to generate any kind of graph. 
\changed{Our results show that the surrogate networks that we generate reflect many of the original networks properties with a high degree of accuracy, not just in terms of specific nodes features, as one might anticipate from the local generating mechanism, but with respect to general features, such as the number of interactions, the number of interacting individuals in time and density of their connections. 
We notice that the characteristics that are better preserved coincide with the ones that depend on time and describe the temporal behavior of the specific nodes, while global features that deal with the spatial organization of the network and which can be observed for instance when collapsing the temporal layers (like the existence of communities) are more difficult to reproduce with this method.}

\changed{In general, this work places itself as a first step towards a deeper understanding of time-varying interaction processes. It allows us in particular to set the spatio-temporal scale of the minimal fundamental knowledge that is necessary to capture many of the intrinsic characteristics that we aim to reproduce in a temporal network.}
The possibility to generate surrogate networks that resemble an original one serves as a test to demonstrate the method efficiency.

\section{Results}
\label{sec:results}
We first briefly sketch the temporal graph generation process. 
Then, we use our method to generate temporal graphs which reproduce the temporal interaction patterns of a diverse set of face-to-face interaction networks, including a hospital~\cite{vanhems2013estimating}, a workplace~\cite{genois2015data}, and a high school~\cite{10.1371/journal.pone.0107878}.
See \textit{Methods} for details on the datasets. 
We evaluate the quality of the generated networks in terms of interaction statistics, considering both static and temporal network properties, highlighting the advantage of our proposed method relative to the state-of-the-art. 
Finally, we show how the approach can be used to expand existing temporal networks, both in time and in number of nodes, something which is not possible using \changed{other} 
methods for surrogate networks.

\subsection{The neighborhood generation process}
\label{sec:method_overview}

\begin{figure}[!h]
\begin{center}
\includegraphics[width=\textwidth]{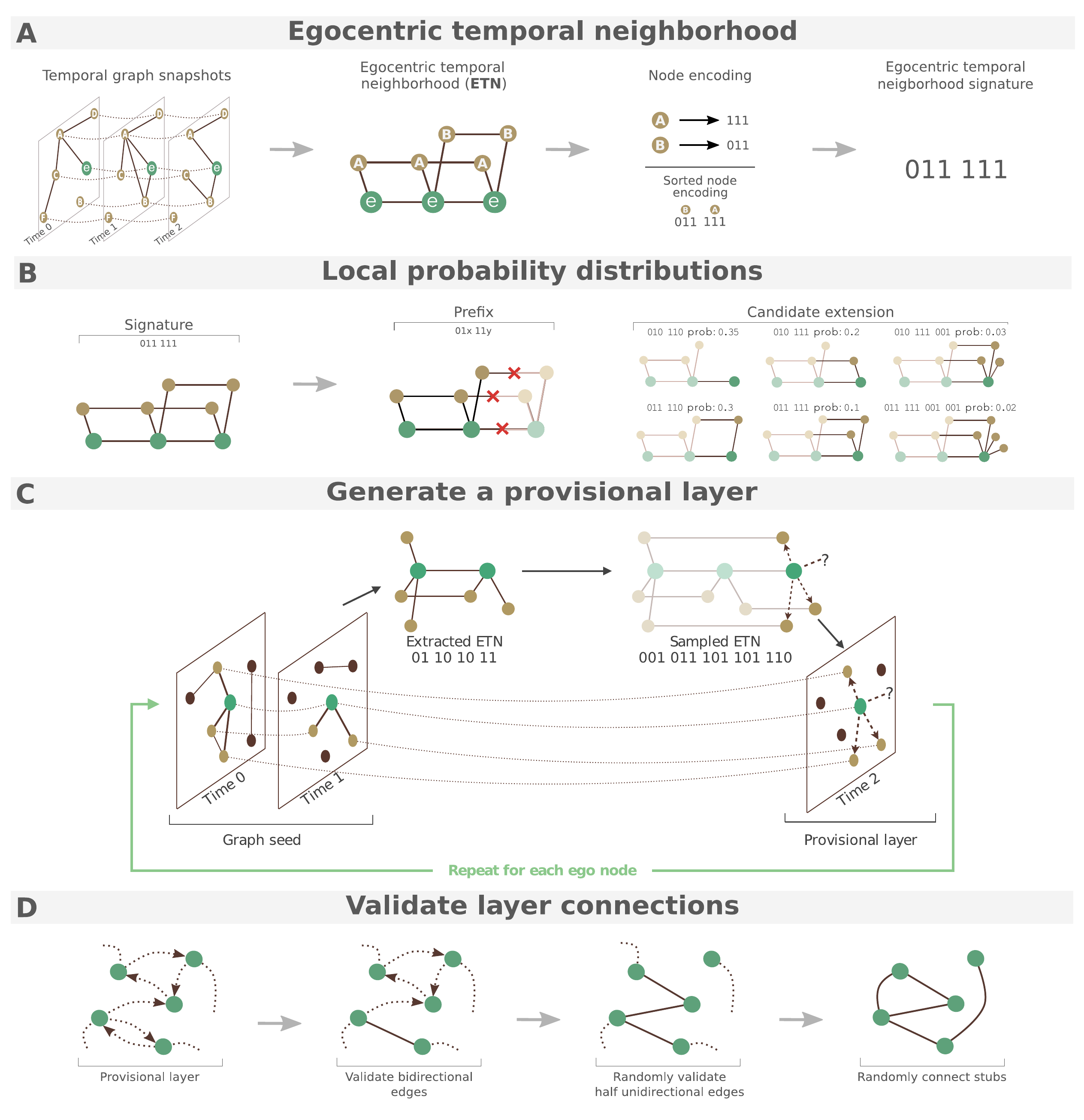}
\caption{\textbf{\etngen.}  
The top panel shows how egocentric temporal neighborhood signatures are extracted and computed. Panel \textbf{B} shows how to build the probability distribution of neighborhoods, necessary to generate a provisional layer. Panel \textbf{C} shows how to generate a provisional layer, while panel \textbf{D} explains how to convert the provisional layer into a definitive one.
}
\label{fig:fig0}
\end{center}
\end{figure}
Figure~\ref{fig:fig0} shows a graphical representation of the generation process for a small temporal network with three timesteps (see \textit{Methods} for details). 

\changed{Our generative algorithm uses as building block the Egocentric Temporal Neighborhood~\cite{longa2021efficient} of a node ($\mathcal{E}_i^{\{t-k,\ldots,t\}}$ for node $i$), which represents the neighborhood of a node over a (short) temporal span $k$. For the sake of compactness, we will refer to the Egocentric Temporal Neighborhood as ETN or simply neighborhood in the rest of the manuscript. Panel A of Figure~\ref{fig:fig0} shows the neighborhood of a specific node, denoted as $e$, for a temporal span $k=2$. $\mathcal{E}_e^{\{t-k,\ldots,t\}}$ contains $e$ and its neighbors at each of $k+1$ consecutive timestamps, discarding connections between the neighbors of $e$, and adding (temporal) connections among instances of the same node at different timestamps.}
Having discarded links between neighbors, $\mathcal{E}_e^{\{t-k,\ldots,t\}}$ can be encoded as a binary string, where for each neighbor node and timestamp \texttt{1} (resp.~\texttt{0}) indicates the presence (resp. absence) of a link connecting to the node at that timestamp. 
Such neighborhoods are extracted for all nodes and all timestamps by using a sliding window over time. 
Notice that a string of 0 and 1 of length $k+1$ is obtained for each neighbor of $e$ but the identity of these nodes is not stored and the final signature does not include any identity labels, just the shape of interactions between neighbors and $e$, as shown in the last step of panel A. This implies that the same specific $\mathcal{E}_i^{\{t-k,\ldots,t\}}$ can be found multiple times in the network, referred to different nodes $i$, different neighbors, and different $t$.

Second (panel B), we build a local probability distribution designed to enable simulation of activity in future time steps.
This distribution to extend the graph into future time steps is based on past neighborhood activity.
Specifically the local distribution maps neighborhoods of length $k-1$ (i.e. temporal neighborhoods involving $k$ steps, \changed{denoted as "Prefix" in the figure}) to the set of all possible extensions into the future (i.e. neighborhoods of temporal depth $k$, involving $k+1$ steps), 
with associated probabilities estimated by Maximum Likelihood over the whole dataset. Basically frequencies of neighborhoods of temporal depth $k$ are first collected from the original temporal network, and then normalized by dividing them by the sum of the frequencies of the neighborhoods sharing the same $k-1$ prefix. \changed{These are denoted as "Candidate extensions" in the figure, where some examples of possible extensions of the prefix are depicted, with their signature and their probability.}

Third (panel C), \changed{we build the surrogate network layer after layer. Given the first $k$ layers, we generate the subsequent one by sampling future interactions for each node from the local probability distribution described above, thus generating a provisional temporal extension of each node. 
Notice that since local probability distributions ignore node identities, future interactions can only involve previously existing neighbors or novel (still unknown) nodes (question mark in panel C).} All the interactions extracted for each node $e$ are represented as directed links from $e$ to its desired neighbors (including ``stubs'', representing links-to-be to unknown nodes).
We thus obtain a provisional \changed{directed} temporal layer of the network.

Last (panel D), this provisional layer is finalized by combining provisional temporal extensions of all nodes, resolving conflicts and dangling links \changed{so as to preserve as much as possible each node's desired neighborhood.} 
We consider a connection from node $i$ to node $j$ in the provisional layer a `request' of $i$ to be connected to $j$. 
If this request is reciprocal the link is validated and added to the new temporal layer (second step in panel D). 
All remaining one-directional links are validated with probability $\alpha= 1/2$ (third step), to preserve the overall number of connections (an $i-j$ connection can be requested by $i$ or by $j$). 
Finally, stubs are pairwise matched up at random (last step in panel D).
The procedure is repeated as many times as the desired length of the final temporal graph, always considering the last $k$ timestamps as seeds to generate an additional one. 

With the basic mechanisms in place, we take a step back and explain how to initialize the process, i.e.~how to obtain the first $k$ layers of the graph. 
The graph at the first timestamp is generated using a configuration model~\cite{molloy1995critical, newman2001random}  reproducing the degree distribution of the first layer of the original graph. 
The following layers up to $k$ are generated by applying the procedure in Figure~\ref{fig:fig0} to the first layer with $k'=1$, to the first two layers with $k'=2$ and so on until $k'=k$.

\changed{Temporal networks are often characterized by an intrinsic periodicity~\cite{holme2012temporal}. This can be captured in our generation process by collecting multiple local probability distributions from the original graph, associated for instance to different days of the week or times of the day.}
%
In the experiments in this paper we use distinct week/weekends or daily local probability distributions, depending on the length and variability of the input network.

The recursive procedure poses no limit to the temporal extension of the network, allowing to generate as many temporal layers as desired, even more than those existing in the original network. Plus, the number of nodes too can be set independently of the original network size (Section~\ref{sec:extension}).  

Above, we have described the simplest possible strategy for extending a layer into the future, but note that all random choices in the link validation process could become preferential choices in order to optimize a specific characteristic of the final network (see Section~\ref{sec:topo_res}).

\subsection{Model evaluation}
\label{sec:outcome}
We now evaluate the quality of the generated networks based on interaction statistics by comparing the networks to empirical data as well as networks generated by a suite of state-of-the-art temporal network generation methods described below. 
We evaluate performance in terms of individual layer topology as well as temporal behavior.

The state-of-the-art methods we consider are:
\dymond~\cite{zeno2021dymond}, a model which  uses the distribution of 3-nodes structures in the original graph (triads with one, two or three connections) as building blocks to generate a new temporal network;
\stm~\cite{purohit2018temporal}, a generative model based on the distribution of small temporal motifs; and 
\taggen~\cite{zhou2020data}, based on deep learning, which uses a generative adversarial network to generate temporal walks that are then combined into a temporal graph.
\dymond and \stm only consider local information, while \taggen is more global.
\changed{These three methods were selected based on the capacity to generate temporal (rather than static) networks, their performance and the implementation availability.}

It is important to underscore that these network generation methods have not necessarily been developed with the aim of generating large temporal networks with low computational cost (see Section~\ref{sec:comp_time}). 
This means that, for example, they require much more training data, need denser temporal snapshots, and therefore \changed{struggle to} generate high temporal resolution networks. 
In particular, both \dymond and \stm require a triad motif to appear in the snapshot. This assumption is too strong for fine-grained snapshots, (i.e. a snapshot every minute). On the other hand, since \taggen is based on a deep learning technique, it requires a massive amount of data in the training phase.
\changed{Differently \etngen, thanks to the linearization allowed by the egocentric perspective, can scale to arbitrarily sized temporal networks.}
In the rest of the paper we report experiments only on the three smallest face-to-face interactions datasets, collected in the hospital~\cite{vanhems2013estimating}, in the workplace~\cite{genois2015data}, and in one of the high schools~\cite{10.1371/journal.pone.0107878} respectively. Results applying \etngen to larger datasets are reported in the Supplementary Information (SI) \changed{and compared only to \taggen due to computational complexity of the other methods.}

\begin{figure}[!ht]
\begin{center}
\includegraphics[width=\textwidth]{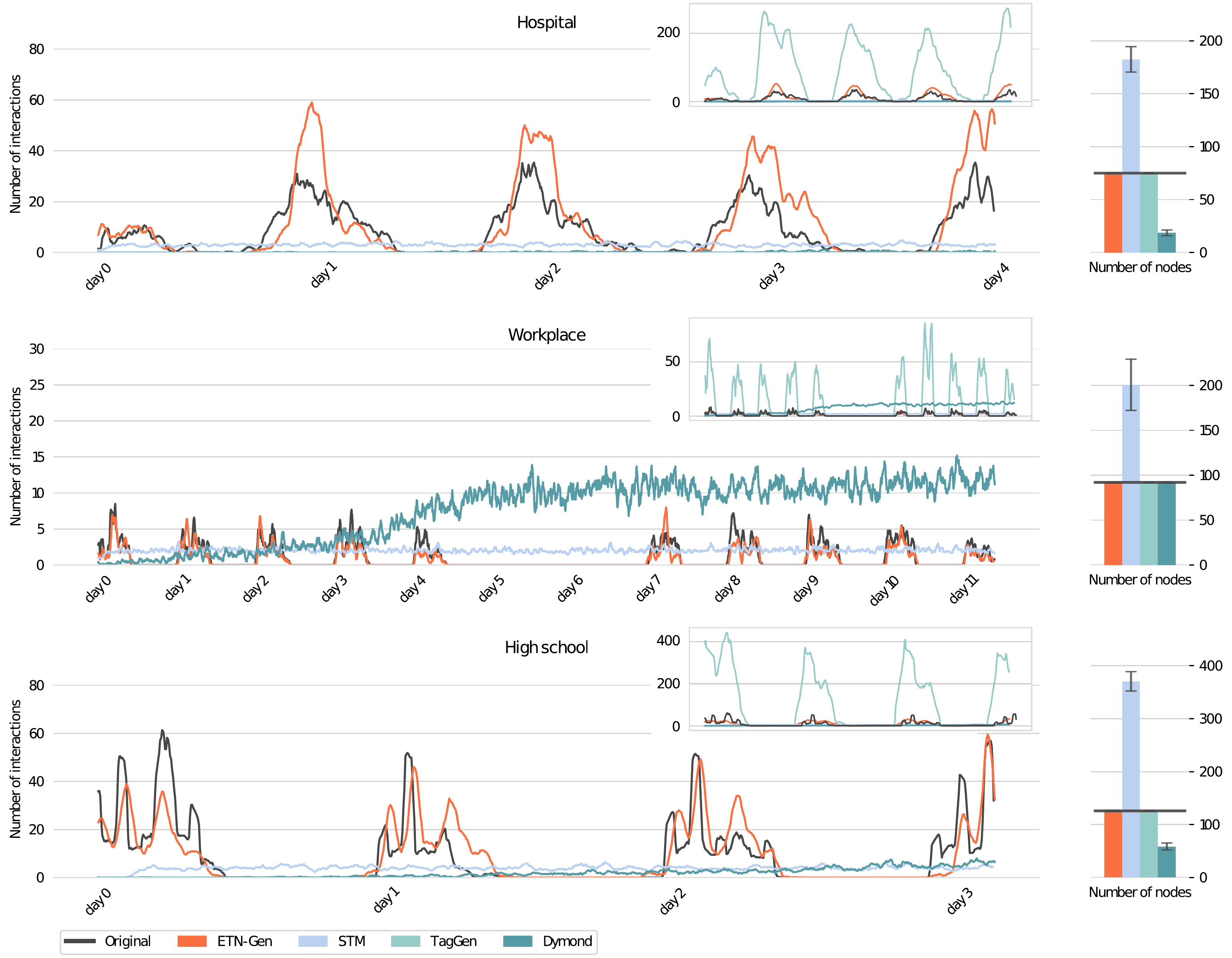}
\caption{\changed{\textbf{Number of interactions in time and number of nodes in the original and surrogate networks.} Each color represents a different generation algorithm, while the original graph is depicted in black. The insets depict the same curves with a different $y$-scale for visibility (the results obtained for \taggen are only reported there).
}
}
\label{fig:nb_inter}
\end{center}
\end{figure}

Figure~\ref{fig:nb_inter} reports the total number of interactions for each temporal snapshot (left) and the average number of nodes (right) in the original network, \etngen and the three \changed{state-of-the-art methods.} 
The first clear finding from this figure is that \etngen (orange curves) results in time-series that are remarkably similar to those appearing in the original datasets (black curves). 
This is true, not just in terms of generating a number of interactions which is of the same order of magnitude as the original data (notice that different datasets have different scales on the $y$-axis), but also in terms of temporal patterns which are preserved with considerable accuracy, including daily and weekly periodicity. 

This result should not come as a surprise, as it is a direct consequence of our network generation procedure. The local probabilistic models store the probability distributions of the neighborhoods appearing in the original graph and this indirectly contains the key information about how nodes degree evolves in time. 
Further, our seed-network has the same degree distribution as the original graph, which allows us to statistically preserve the overall average number of interactions of the original graph. 
Moreover, we manually input periodicity via different local probabilistic models for different times and days of the week. 
We highlight, however, that while using only a single local probabilistic model would remove our ability to model periodic changes in graph over time, we would still be able to model the average number of interactions, as these are automatically reproduced by the rest of the algorithm. A detailed analysis is reported in the SI (Section \ref{si:multiple_dict}).

\changed{From the comparison with the other methods we conclude that \etngen is the only method able to preserve the number of nodes, the order of magnitude of the amount of interactions and the periodicity of the original network.}
The curves for the original network and \etngen are also reported in Figure~\ref{fig:nb_inter_generalization_notag} in SI for larger datasets to which the other methods cannot be applied due to computational constraints.

\begin{figure}[!ht]
\begin{center}
\includegraphics[width=0.95\textwidth]{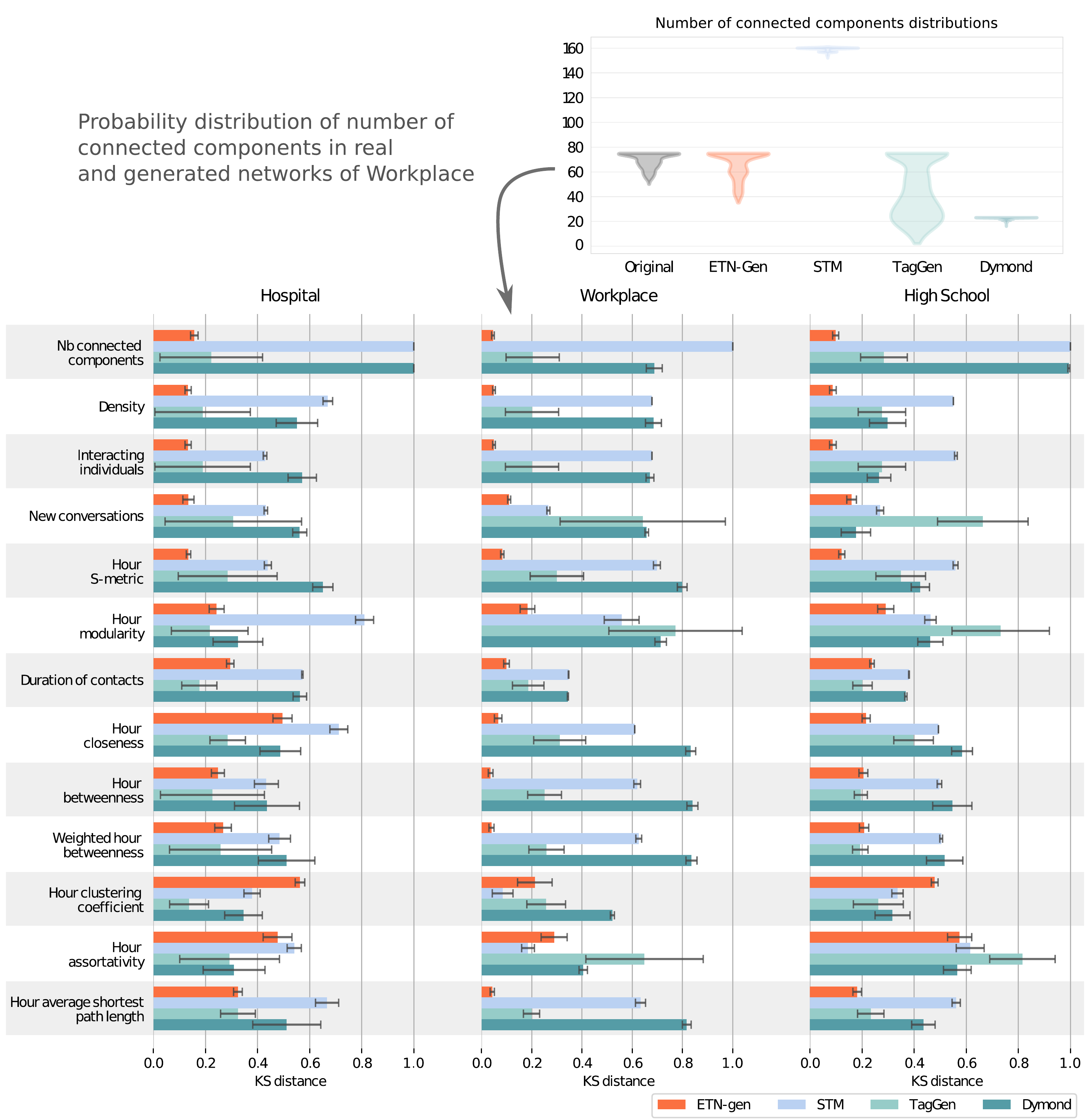}
\caption{\textbf{Topological similarity according to time-dependent measures.} Similarity of the original network with those generated by \etngen, \stm, \taggen and \dymond. Each bar reports the Kolmogorov-Smirnov distance between the two distributions (original and generated) for a specific structural metric. The shorter is a bar the more similar are the distributions. Standard deviations are obtained over 10 stochastic realizations of each network. In the top inset we report the distributions of the \changed{number of connected components} in real and in one instance of generated networks for the Workplace dataset.}
\label{fig:top_kstest}
\end{center}
\end{figure} 

\begin{figure}[!ht]
\begin{center}
\includegraphics[width=\textwidth]{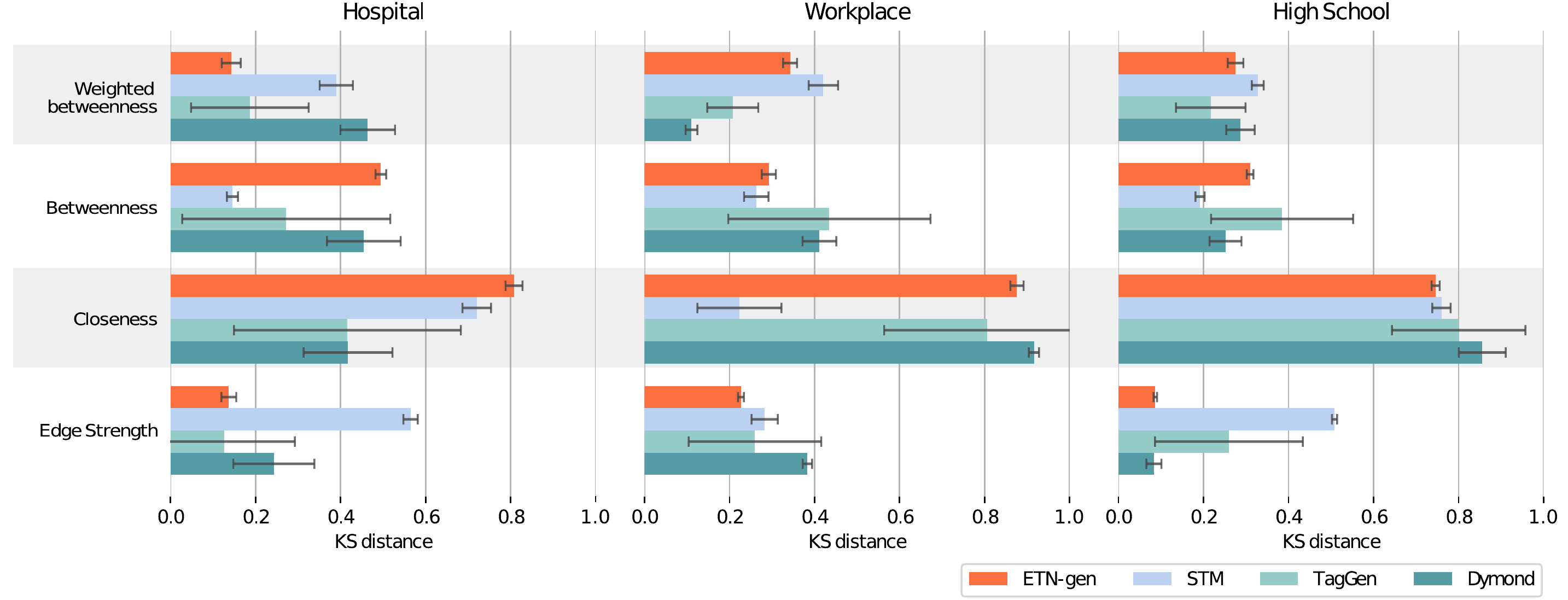}
\caption{\changed{\textbf{Topological similarity according to time-aggregated measures.} Analogous to Figure~\ref{fig:top_kstest} for measures on the aggregated networks.}}
\label{fig:top_kstest_timeaggregated}
\end{center}
\end{figure}

\subsection{Topological similarity evaluation}
\label{sec:topo_res}
Having studied the temporal development, we now turn to structural similarity between the surrogate data and the original networks. 
We consider \changed{seventeen} metrics for structural similarity, \changed{divided between those that depend on time, namely: number of connected components~\cite{newman2018networks}, density~\cite{zeno2021dymond}, number of interacting individuals~\cite{starnini2012random}, new conversations~\cite{starnini2012random}, hour S-metric~\cite{li2005towards}, hour modularity~\cite{blondel2008fast,clauset2004finding}, duration of contacts~\cite{starnini2012random}, closeness~\cite{newman2018networks}, hour betweenness (weighted and unweighted)~\cite{newman2018networks}, hour clustering~\cite{luce1949method, wasserman1994social}, hour assortativity~\cite{newman2002assortative}, hour average shortest path length~\cite{holme2012temporal}; and those that are measured on the aggregated network (i.e. collapsing all the temporal layers in one weighted network), which are: closeness~\cite{newman2018networks}, betweenness (weighted and unweighted)~\cite{newman2018networks}, and edge strength~\cite{starnini2012random}.
All the measures are collected as distributions, the temporal ones measured on each singular temporal layer or on slots of one hour (those whose names start with "Hour"), and the aggregated ones as distributions over the edges or the nodes. See SI for the definition of all measures.
}

To compare distributions we rely, inspired by Zeno et al.~\cite{zeno2021dymond}, on the Kolmogorov-Smirnov distance~\cite{massey1951kolmogorov} to contrast generated and original graphs. \changed{In the SI, we also consider alternative distance measures, namely the Jensen-Shannon divergence~\cite{lin1991}, the Kullback-Leibler divergence~\cite{kullback1951}, and the Earth mover's distance~\cite{mallows1972}, obtaining similar results (see Section~\ref{sec:si:otherDist} in SI).} 
Distances between distributions are reported in Figures~\ref{fig:top_kstest} and \ref{fig:top_kstest_timeaggregated}, where we compare graphs obtained with \etngen with those from the three alternative approaches. 
The networks generated with \etngen (orange bars) show a high similarity to the original networks for many of the measures and also a high stability (small errorbars).
The measures for which \etngen performs best are those that, together with the number of interactions (see Figure~\ref{fig:nb_inter}), are preserved by construction: the density and the number of interacting individuals in time. 
Here, the similarity originates from the neighborhood probability distributions, which ensure that from a statistical viewpoint, the surrogate network has the same number of interactions and the same number of individuals involved in an interaction.
The same holds for the number of times that a new link appears, as these statistics are also stored in the neighborhood probability distributions.
Another characteristic that is 
\changed{well} captured by the egocentric temporal neighborhoods is the hub-like structure that we can find in each static layer, which is measured by the S-metric~\cite{li2005towards}.

Going beyond these `trivial' consequences of the mechanics of the generating mechanisms, the method does well at preserving \changed{the number of connected components. Indeed, the inset shows how the \etngen networks exhibit a distribution of the number of connected components that is similar to the original one, while \taggen shows a rather larger distribution difference and the other methods generate substantially fewer (\dymond) or more (\stm) connected components. 
This is a consequence of the fine-grained temporal information that \etngen uses to generate the networks. For the same reason, the hour modularity of \etngen networks is always better or comparable with that of the other generated networks.
}

\changed{The distribution of durations is instead not very well reproduced, but it is not bad with respect to the other methods. In fact, considering a $k$-steps memory allows interactions to have a continuity in time, differently from the case of independent layers. 
} 

\changed{We also test three different centrality measures. Since centrality is a quantity that characterizes each node at each time, we focus first on the temporal distributions, by reporting for each temporal slot lasting one hour the nodes average (see Figure~\ref{fig:top_kstest}). Then we consider the spatial distribution, reporting the centrality of each node computed on the  time aggregated network (see Figure~\ref{fig:top_kstest_timeaggregated}). We observe that when we consider the temporal distribution we obtain a higher similarity to the original networks, confirming the insight that \etngen is more valid in reproducing fine-grained time features, while it results more limited in reproducing the global spatial organization of the networks.} 

Another interesting property is the distribution of edge strengths in the projected graph (see Figure~\ref{fig:top_kstest_timeaggregated}). 
Edge strength is simply the number of times that each edge has appeared over the duration of the graph. 
Here, we would not necessarily expect \etngen to do well as the method will tend to create networks with quite homogeneous distributions of strength. 
This is because it can only rely on a memory of order $k$ for edge repetitions, and does not have a long-term memory.
Hence all the heterogeneous behaviors that we can find for instance in social datasets, where individuals tend to establish relationships with specific nodes and have repeated (but not necessarily consecutive) interactions  with them, are not preserved by \etngen. 
Nevertheless, we find that for the considered datasets \etngen remains competitive with the other methods.

If edge strength is partially affected by the absence of long memory, the most important limitations of the egocentric perspective are highlighted by clustering, degree assortativity and average shortest path length, which are related to second-order interactions (see Figure~\ref{fig:top_kstest}). 
This is the cost we pay for having a computationally efficient model applicable to arbitrary networks. 
Notice that while this is a problem in theory, it seems not to affect the workplace dataset, which is a substantially sparser network with low clustering and short paths.


\changed{In general from the topological analysis we observe that the features that are more preserved are the time-dependent ones (at least those that do not depend on second-order interactions), while the method is more limited in reproducing the time-aggregated measures.
This is valid for both local and global features. 
An additional analysis on meso-scale structures is reported in Section~\ref{sec:si:mesoscale} of SI. 
We observe that small static motifs are well preserved by \etngen (Section~\ref{sec:si:motifs}) but not the network communities (if present in the original graph). This is a common limitation involving the other generation methods, too. }


\subsection{Dynamical similarity evaluation}\label{sec:dynamic}
Having tested our method from the structural point of view, we now test the usefulness of the surrogate networks in terms of dynamical processes unfolding upon them. 
We study two dynamical models: random walk and a spreading model.

\begin{figure}[!h]
\begin{center}
\includegraphics[width=\textwidth]{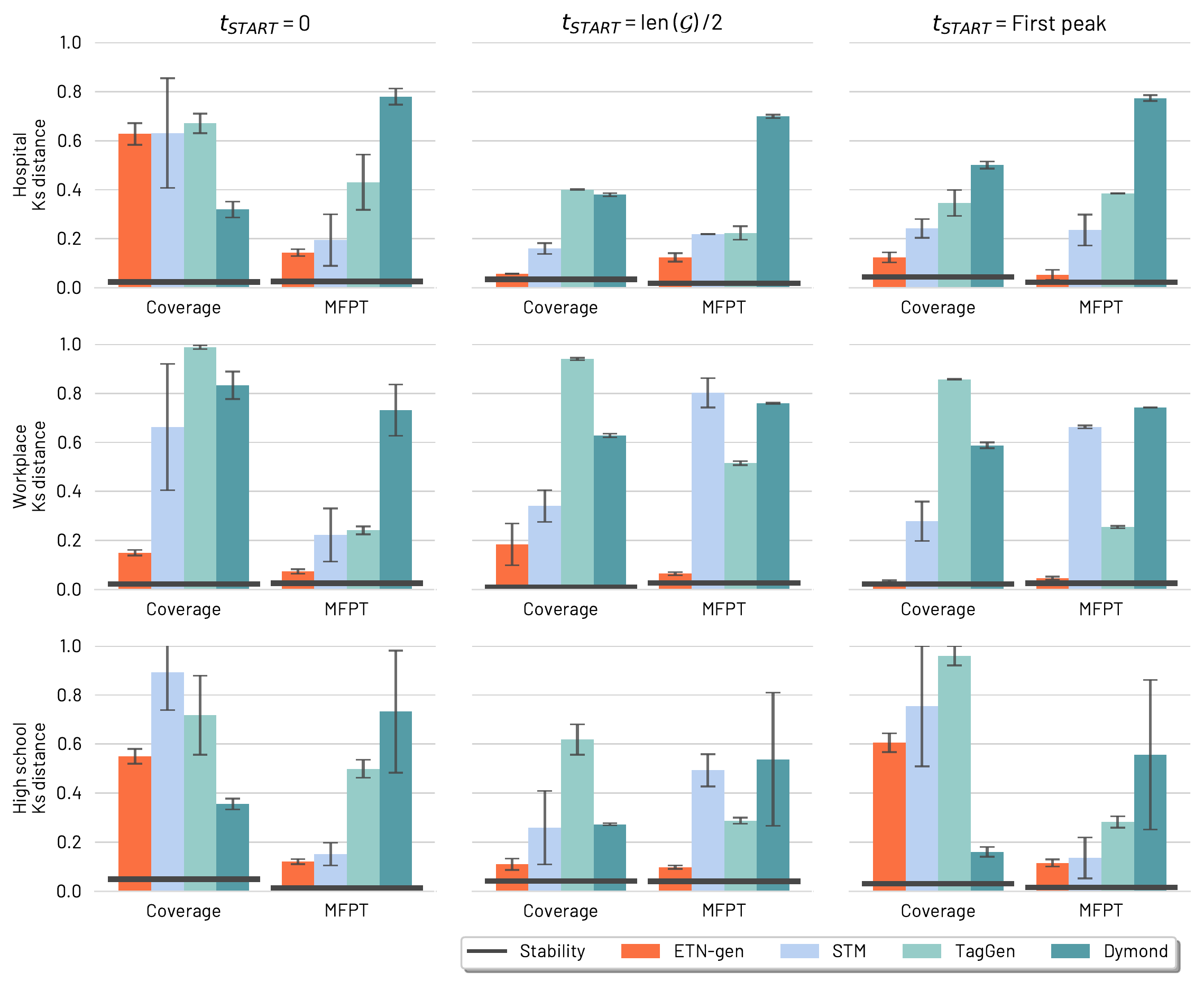}
\caption{\changed{\textbf{Dynamic similarity: Random walk.} Kolmogorov-Smirnov distance between original and generated distributions of coverage and mean first passage time in the random walk model in each generated network for three different starting points: time 0, T/2 and on the first peak. Our method is represented in orange, while the solid black line shows the stability (i.e. the same simulation on the original network).}}
\label{fig:rw}
\end{center}
\end{figure}

\subsubsection{Random walk}
We simulate a temporal random walk~\cite{starnini2012random, holme2015modern} on the original and generated networks. We use the standard definition of random walk extended to temporal networks: a random walker starts from a randomly chosen node at \changed{generic time $t$} and chooses uniformly at random one of its neighbors, moving there. Then the second step will take place on the following layer  of the temporal network, so the walker will randomly choose between its neighbors at time $t+1$, and so on, assuming that at each time corresponds one and only one jump.

We compute two metrics: coverage and mean first passage time (MFPT), and compare distributions over different realizations between the input and the generated temporal network using again the Kolmogorov-Smirnov distance (see SI for the definition of the metrics). \changed{We consider three different starting points: $t=0$, $t=len(G)/2$ (in the middle point of the temporal extension of the graph), and the time corresponding to the first peak of  connections (when the number of connections reaches a maximum). }


In Figure~\ref{fig:rw} we report the Kolmogorov-Smirnov distance for coverage and MFPT. 
The horizontal dashed line shows the stability of each measure on the original network. 
The black line is obtained comparing different performances (average over 1000 simulations of random walk for coverage, and 5 times each couple of nodes for mean first passage time) by means of the Kolmogorov-Smirnov distance. \changed{It is worth noting that the stability is different from zero, due to inherent variations within the dynamic process.}
\changed{We observe that the dynamics on the \etngen networks are similar to the ones on the original networks in terms of mean first passage time, while in terms of coverage performance they depend on the datasets and the starting point but they always results competitive with the ones obtained with the alternative methods.}
\changed{In Section~\ref{sec:si:dyn} we also report the evolution in time of the number of newly visited nodes.}

In general we can say that the random walk process on the \etngen's surrogate networks is quite similar to the random walk on the original graph. 

\begin{figure}[!h]
\begin{center}
\includegraphics[width=\textwidth]{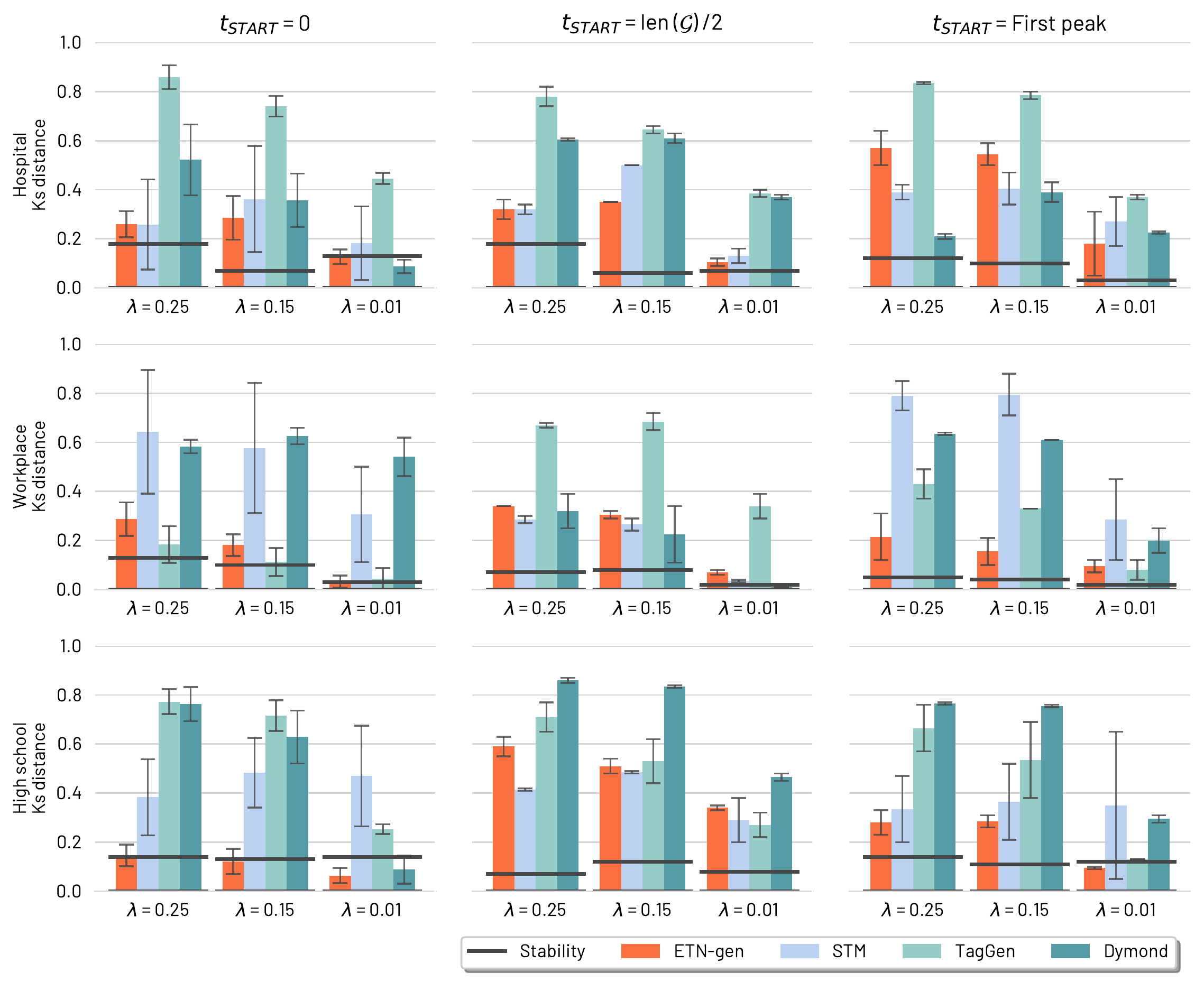}
\caption{\changed{\textbf{Dynamic similarity: Spreading model.} Kolmogorov-Smirnov distance between original and generated distributions of $R_0$ values on a SIR model simulation in each generated network for three different starting points: time 0, T/2 and on the first peak. Our method is represented in orange, while the solid black line shows the stability (i.e. the same simulation on the original network).}}
\label{fig:sir}
\end{center}
\end{figure}

\subsubsection{Spreading model}
We simulate a Susceptible-Infectious-Recovered (SIR) model~\cite{may1991}, with three possible values for the probability of disease transmission ($\lambda \in \{0.25, 0.13, 0.01\}$), and the recovery rate fixed at $\mu=0.055$.  \changed{In each simulation the infection starts at time $t_{START}$ by assigning to one random node (selected among the connected ones at that time) the status of infected.}
\changed{This initial node will infect its neighbors with probability $\lambda$ and recover with probability $\mu$. In the next time step we consider the following temporal layer of the  network and again the infected nodes can infect their new neighbors or they can recover. We repeat the procedure until the end of the temporal layers or until all the infected nodes have recovered. Again, we consider the three different starting points described above also for the dynamics.}
We compute the reproduction value $R_0$ (see SI for the definition).
Each experiment was repeated 100 times and the distribution of $R_0$ obtained on the original network is, again, compared with those obtained on synthetic networks by means of the Kolmogorov-Smirnov distance. 
Results are shown in Figure~\ref{fig:sir}, where again a horizontal black line shows the stability of each measure on the original network (computed averaging over 100 simulations). 
We observe that the results obtained with \etngen are highly similar to those of the original graph and show a large degree of stability with respect to this similarity. 
\changed{In Section~\ref{sec:si:dyn} we also report the evolution in time of the number of infected nodes.}

\subsection{Dataset expansion and extension}
\label{sec:extension}

In the previous sections we have argued that \etngen creates realistic surrogate temporal networks that mimic \changed{many aspects of} real social dynamics (both in terms of structure and in reproducing dynamical systems).

Now we ask the question: How can this tool be useful in practice? 
A relevant application is represented by the possibility of enlarging a given temporal dataset, both in time and in size. 
It is indeed common that a specific analysis, in order to yield reliable results, requires a larger population or a longer time than those characterizing collected real data. 
In those cases we deal with the long-standing problem of data augmentation, for which we now argue that \etngen represents a promising solution. 
In the following we show how our method can be used for augmenting a temporal dataset, by adding temporal layers (temporal extension), but also by increasing the size of the network in terms of number of nodes (size expansion).

\subsubsection{Temporal extension}
\begin{figure}[!h]
\begin{center}
\includegraphics[width=\linewidth]{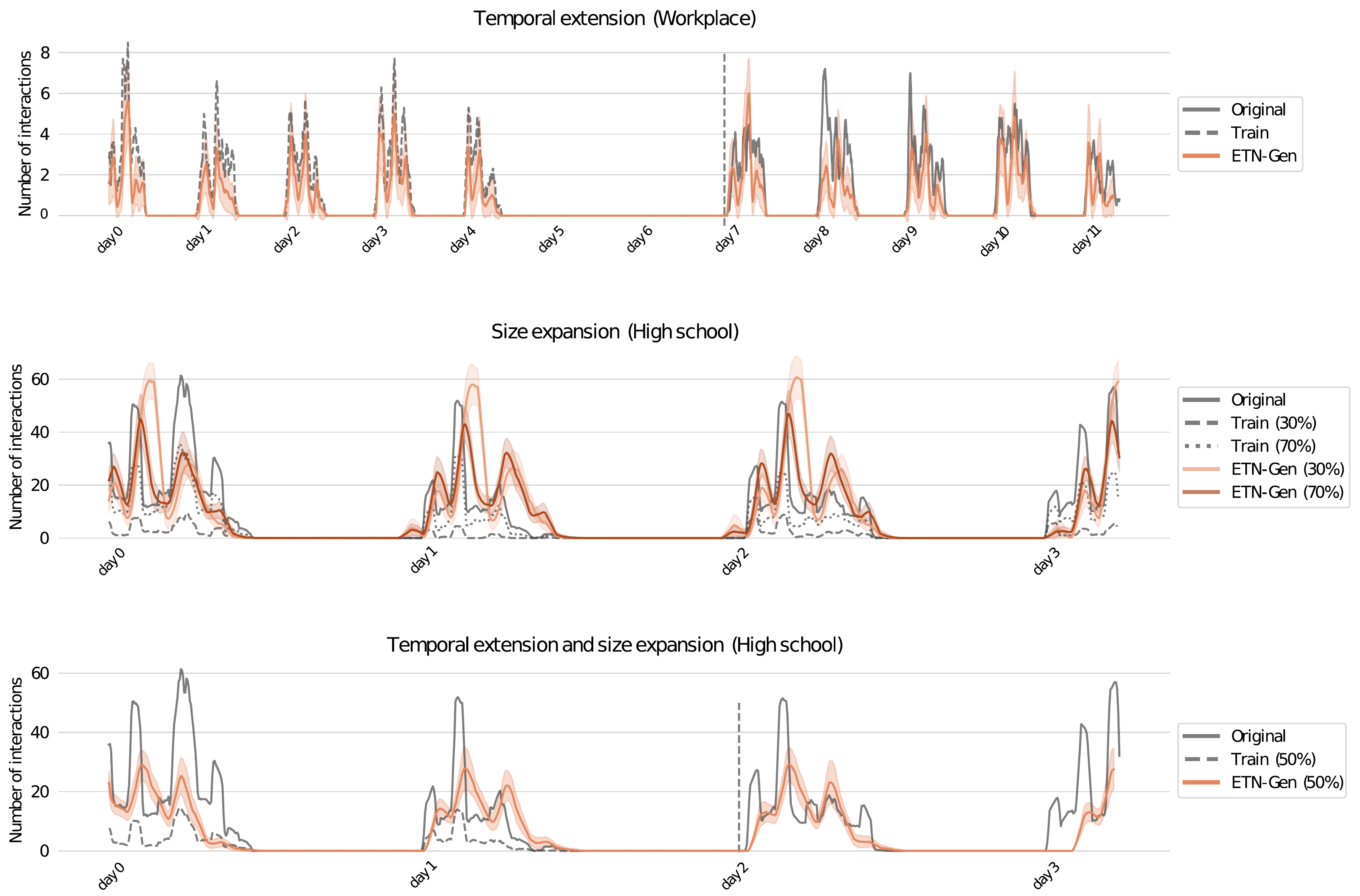}
\caption{\textbf{Temporal extension and node expansion.} 
The mean and standard deviation of our method are shown in orange (and brown).
Black dashed (and dotted) lines show the original data used to train our model, while black solid lines show the original data used to evaluate the quality of the generated network.
In experiments involving temporal expansion, a vertical bar separates the temporal range used to collect training data from the one where expansion is performed. 
}
\label{fig:moreLength_moreNode_moreLengthNode}
\end{center}
\end{figure}

The procedure, as explained in the previous sections, implies calculating the neighborhood probability distributions, which somehow summarizes the interaction patterns in the original graph. Each layer of the surrogate networks is built extracting possible interactions for each node from these distributions, a process that only depends on the last $k$ layers. The temporal extension of a dataset is therefore straightforward: the procedure of temporal layer addition can be repeated possibly an infinite number of times, and we stop when the desired number of time steps is reached.
At the top of Figure~\ref{fig:moreLength_moreNode_moreLengthNode} we show an example of temporal extension of the workplace network. 
We have selected this dataset to highlight the ability of \etngen to differentiate between week days and weekends. 
To evaluate the quality of the extension, we assume to only know the first week of the original two-week dataset (from the beginning to the vertical line) and from this we estimate the neighborhood probability distributions. 
We then use it to generate an ensemble of 10 surrogate networks with a length of two weeks. 
The mean and standard deviation of the number of interactions in the generated graph are reported in orange. 
The number of interactions of the real graph are reported in black dashed curves for the first week, and in black solid curves for the following week. In other words, the method is  ``trained'' on the first week of the real dataset, several two-week networks are generated, and they are eventually ``tested'' comparing them with the two-weeks-long real dataset.
Results show how the generated networks accurately recreate the original behavior beyond the timespan that was used to estimate the local probability distributions.

\subsubsection{Size expansion}

Here we explore the fidelity of surrogate networks with an increased number of nodes. 
As discussed above, it is possible to increase the size beyond that of the original network within the \etngen framework because the number of nodes is simply a parameter to set for the method. 
That said, however, the concept of size expansion requires more attention than time extension. 
Because, as we change the number of nodes in a network we should also consider how the density of the graph and the mean degree should change accordingly. 

In the following we describe an experiment of data augmentation, assuming that we only have access to incomplete data.
Incomplete data are obtained by randomly removing part of the nodes from the original network. 
We use the high school dataset which, with its 126 nodes, is the largest among our datasets, and we consider two reduced versions, with 30\% and 70\% of the nodes respectively. 
When removing part of the nodes from a network, we naturally remove also part of the links (all those which were before connecting the eliminated nodes to the remaining ones), we hence reduce the mean degree. 
We should consider that an incomplete dataset has in general a reduced mean degree with respect to the real-world network, and that when we try to reconstruct the original network via data augmentation we should increase the mean degree too. 
See Methods for a quantification of the needed increase.

Anyway, once the desired connectivity has been chosen, \etngen allows us to generate a surrogate network with the desired number of nodes and the desired degree, while maintaining the pattern of egocentric interactions of the original dataset.

The results of the experiment on the high school dataset are shown in the middle panel of Figure~\ref{fig:moreLength_moreNode_moreLengthNode}.
For each of the two reduced temporal networks we generate a temporal network with 126 nodes to try to reconstruct the original graph. 
We generate the initial snapshot using the configuration model based on the degree distribution of the first snapshot of the original (not reduced) graph. 
Then we build local probability distributions only using information from the reduced networks and use these local probability distributions to generate surrogate expanded networks from them. 
The expanded networks have the same number of nodes of the original one (126), enabling direct comparison. The expedient that we use to augment the mean degree from the reduced seed graph is to increase the parameter $\alpha$ of the generation process, which is the probability to confirm the uni-directional directed links in each provisional layer (set to $1/2$ by default). See Methods for the details on how to compute the correct value of $\alpha$ given the original number of links and the desired density of the generated graph.

In the middle panel of Figure~\ref{fig:moreLength_moreNode_moreLengthNode} the black solid curve represents the number of interactions in the original network, the black dashed curve those in the ``train'' network with 30\% of the nodes and the black dotted curve those in the one with 70\% of the nodes. 
The corresponding values for the generated networks with their standard deviations are reported in orange and brown respectively. 
Again, we observe the ability of our method to correctly replicate the pattern of interaction in the original network, even if fed with a small percentage of nodes from the original graph as seed.

\subsubsection{Temporal extension and size expansion}
We can also combine the two techniques above to simultaneously increase the number of nodes and the temporal snapshots. 
The results are shown in bottom panel of Figure~\ref{fig:moreLength_moreNode_moreLengthNode} for the high school network, where the synthetic graph has been obtained by only using 50\% of the nodes and the first two days of the original dataset (from the beginning to the vertical line), see the black dashed curve.
Also in this case, our method is able to extend an input graph in both the temporal and the node size dimensions with remarkable accuracy.

\section{Discussion}
\label{sec:discussion}
In this manuscript we have proposed a model to generate surrogate temporal networks, i.e.~synthetic networks that realistically capture \changed{many} properties of real-world datasets, only making use of the information contained in egocentric temporal neighborhoods. 
Specifically, we generate temporal networks which accurately reproduce structural characteristics like density,  number of interacting individuals, \changed{number of connected components,} and the possible presence of hubs. 
We observe that in both topological and dynamical tests, the networks generated by this model are generally closer to the original graph than those generated by different literature models. 
Moreover, this approach is able to generate temporal networks that have different sizes than the original one. 
This property can be used to increase the number of nodes and extend the network in time, providing a powerful tool for data augmentation.
\changed{
These results suggest that egocentric temporal neighborhoods, that we use as building blocks, contain fundamental information about the real networks they are extracted from.}

By using \etngen surrogate networks it is possible to overcome privacy issues, too. We did not explicitly prove that such surrogate networks are impossible to de-anonymize, but we are rather confident in the privacy-preserving properties of the method. In fact, the interactions of one node in the surrogate are designed based on the probability distribution of ETN prolongation, that is in its turn constructed based on the interactions of all the nodes in the original graph (remembering that the identity of nodes are not stored). Therefore there is not a match node-to-node between surrogate and original graph. More precisely, the set of interactions of one node in the original graph is distributed among multiple nodes in the surrogate graph. We hence find it unlikely that real nodes can be reconstructed and identified observing the surrogate network. \changed{However, this will be matter of future investigations.}

The other side of the coin is that this simplicity does not capture certain topological features. 
This is the main limitation of the model. 
For instance, disregarding second-order interactions translates to a reduced ability to preserving clustering, degree correlations and average shortest path length. \changed{This is the price to pay to achieve scalability, sidestepping the graph isomorphism problem in mining egocentric temporal neighborhoods. Just getting rid of this simplifying assumption would imply a substantial blow-up in computational complexity (as suggested by the runtime comparisons with alternative approaches reported in Table S1) and, as a consequence, a significant reduction in the timespan of the temporal neighborhood that could be dealt with. Trading second order interactions for longer temporal neighborhoods allows us to reproduce most of the relevant features of temporal networks while maintaining computational efficiency. Nevertheless, further research is needed to explore alternative trade-offs in the expressivity-efficiency scale. Another limitation of the proposed approach is the absence of long-term memory, which implies that the model cannot capture long-term patterns of interaction (like e.g. daily or weekly recurrences).} 
These features are instead well captured by more theoretical models of network generation that include aging~\cite{moinet2015burstiness, moinet2016aging}, edge reinforcement~\cite{gelardi2021temporal,stehle2010dynamical}, or in general some mechanism for memory such that contact duration and inter-event times are heterogeneous and depend on the past interactions~\cite{rocha2013bursts,vestergaard2014memory}. 
Memory could also be used to generate a synthetic temporal network that is organized in communities~\cite{zhao2011social, zhang2017random}. 
This is a characteristic often occurring in social networks (particularly evident in schools), and it cannot be captured by small local subnetworks like egocentric temporal neighborhoods. 
However, long-term memory appears in literature only in theoretical models for temporal network generation, for which the goal is to obtain realistic networks by recovering some particular characteristics of the observed dynamics in real networks, but usually do not aim at reconstructing specific real networks or environments. \changed{Indeed, the alternative methods we evaluated in this manuscript also fail to account for long-term memory.}
A model which instead is built to obtain surrogate networks with an alternative approach is the one proposed by Presigny et al.~\cite{presigny2021building}. This model does not generate a new network from scratch, it instead individuates a backbone of a real temporal network, defined as the global subnetwork composed of the most significant edges, and then reconstructs the missing links.
This is based on a conceptually different idea, assuming that the important information concerns the global structure of the network, while the method that we are proposing focuses on how nodes behave given their interactions in last time steps.
\changed{This is indeed evident from Figures \ref{fig:top_kstest} and \ref{fig:top_kstest_timeaggregated}: if \etngen is highly effective in reproducing the evolution in time of singular node neighborhoods, it fails in reproducing global network features that are not reproducible by only using ego-node information.}
By recalling  two different long-standing traditions in network science, a socio-centric versus an ego-centric perspective~\cite{wasserman1994social}, we can assert that if the first one is covered, for what concerns surrogate temporal networks, by the model of Presigny et al.~\cite{presigny2021building}, our model  places itself in the remaining gap, filling the unexplored case of the ego-centric perspective.

The insertion of memory or second order mechanisms, implying the possibility to reconstruct an organization in groups of nodes, and also to make the set of nodes change in time, inserting new nodes or excluding old ones, are demanded to future work, aiming at improving the current method for a further advance in realistic reconstruction.


\section{Methods}
\label{sec:method}

\textbf{Data description and processing.} The three temporal networks studied in the main body of this work represent face-to-face human interactions collected by the SocioPatterns project\footnote{http://www.sociopatterns.org/}: 

\begin{itemize}
    \item \textbf{Hospital~\cite{vanhems2013estimating}.} The dataset has been collected in the geriatric ward of a university hospital~\cite{vanhems2011risk} in Lyon, France, over four days in December 2010. It contains interactions among medical doctors, paramedical staff, administrative staff and patients.
    Number of edges: $1139$, number of nodes: $75$.
    \item \textbf{Workplace~\cite{genois2015data}.} The dataset has been collected in 2013 at the \textit{Institut National de Veille Sanitaire}, a health research institute near Paris, over two weeks. It contains interactions among individuals from five departments. 
    Number of edges: $755$, number of nodes: $92$.
    \item \textbf{High school~\cite{10.1371/journal.pone.0107878}.} The dataset has been collected in 2011 in Lycée Thiers, Marseilles, France, over four days (Tuesday to Friday). It contains interactions among 118 students and 8 teachers in three different high school classes.  Number of edges: $1709$, number of nodes: $126$.
\end{itemize}

As stated by the researchers involved in the aforementioned studies, each study participant and staff member was asked to sign an informed consent and each study received the approval by the French national body responsible for ethics and privacy, namely the “Commission Nationale de l’Informatique et des Libertés” (CNIL, http://www.cnil.fr). More details can be found in the publications describing the studies and the collected data ~\cite{vanhems2013estimating,genois2015data,10.1371/journal.pone.0107878}.

\changed{\textbf{Kolmogorov-Smirnov distance.}
The (two-sample) Kolmogorov-Smirnov test\cite{massey1951kolmogorov} is a non-parametric test used to test how likely it is that two sets of samples come from the same (unknown) distribution. The test uses the following statistic: 
}
\begin{equation*}
    D_{KS} = \max_x |F_1(x) - F_2(x)|
\end{equation*}
\changed{
Where $F_1(x)$ and $F_2(x)$ are the empirical cumulative distributions of the two sets.\\
While originally conceived for hypothesis testing, the KS statistic has often been used to measure the distance between empirical cumulative distributions  \cite{zeno2021dymond,swiderski2015texture,baselice2019denoising,zierk2020reference,lopes2007two,luiz2021application}. We follow this common practise in this manuscript.
}

\textbf{Neighbourhood generation process: parameters.} The gap between two consecutive temporal snapshots has been set to $5$ minutes for face-to-face interaction networks and $10$ minutes for SMS and phone call networks (in SI). The time horizon $k$ defining the egocentric temporal neighbourhood has been set to $k=2$ in all experiments, which is the minimal horizon that preserves some temporal correlation. In section \ref{si:varing K} we motivate our decision in using $k=2$ \changed{but also show the results for $k=3$}. Local probability models have a granularity of 1 hour and a periodicity of 1 day (i.e. between 8 and 9 am in each day we use the same probability model, and the same holds for all 1 hour slots in the day), for all networks but the ones including weekends, namely Workplace and High school 2, for which the periodicity is set to 1 week. 

\textbf{Space and time complexity.} The time complexity required by our method is $\mathcal{O}(n \cdot m)$, where $n$ is the number of nodes in the temporal graph and $m$ is the number of timestamps. The space complexity is constant with respect to both network size and number of timestamps. See Section \ref{si:complex} of  SI.

\textbf{Size expansion: preserving interaction density.} The seed graphs for the size expansion experiment are generated by artificially reducing the original dataset (so that the original graph can be used as ground-truth). In this reduction process, whenever a node is dropped all its connections are dropped too. As a consequence, the resulting seed graph has a reduced mean degree with respect to the original one, and the expanded graph generated from it would inherit this reduced mean degree. This problem can be avoided by adjusting the $\alpha$ parameter of the generation process (the probability to confirm the unidirectional links in each provisional layer, set to 1/2 by default).
In particular, we would need to set $\alpha = 1 - \frac{1}{2} \frac{\hat L}{L}$, where $\hat L$ is the average number of links in the seed graph and $L$ the desired number of links in the generated graph. However, $L$ is unknown and needs to be estimated.
Something that we know, and that we want in this case to preserve, is the density, defined as $d =  \frac{\hat  L }{\hat N\cdot(\hat N-1)/2}$ i.e.  the fraction between the number of links in the seed graph and all possible links ($\hat N$ is the number of nodes in the seed graph). If we assume a linear growth with respect to the number of all possible edges in the network, we also have: $d =  \frac{ L }{N\cdot( N-1)/2}$, with $ N$ as the number of nodes of the generated graph (that we can choose). Combining these two equations we obtain an estimate for $L$, from which we obtain: $\alpha = 1- \dfrac{\hat N  \cdot (\hat N -1) }{N \cdot (N-1)} \cdot \frac{1}{2}$. Hence, when we consider a seed with only 30\% of the nodes of the high school dataset (so $N=126$ and $\hat N= 38$) we should use $\alpha = 0.96$ to reproduce the same density. While if we start with 50\% and 70\% of the nodes (i.e. $\hat N=63$ and $\hat N=88$) in the seed we should use respectively $\alpha = 0.88$ and $0.76$. 

\textbf{Alternatives approaches for generating networks.} \dymond~\cite{zeno2021dymond} builds a temporal network considering (i) the dynamics of temporal motifs in the graph and (ii) the roles nodes play in motifs (e.g. in a wedge -- two links connecting three nodes -- one node plays the hub, while the remaining two act as spokes). The method has no parameters to be set. \textit{Structural Temporal Modeling (\stm)}~\cite{purohit2018temporal} extracts counts for a predefined library of (non-egocentric) temporal motifs from the original network, and turns them into generation probabilities from which to create the temporal network. \changed{In particular, we use the parameterized version of STM and we set the parameter $\alpha = 0.6$ as recommended by~\cite{purohit2018temporal}}. \taggen~\cite{zhou2020data} is a neural-network based approach that extracts temporal random walks from the original graph and feeds them to an assembling module for generating temporal networks. \taggen has been trained with the parameters used in the original paper, namely 30 epochs with a batch size of $64$ and stochastic gradient descent with a learning rate of $0.001$.

\section{Code and data availability}
The data used to support this study are publicly available at the following links. 
\begin{itemize}
    \item The SocioPatterns data at \url{http://www.sociopatterns.org} 
    \item The CNS data at \url{https://doi.org/10.6084/m9.figshare.7267433}
    \item The Friends and Family data at \url{http://realitycommons.media.mit.edu/friendsdataset.html}
\end{itemize}

The codes used for the generation of temporal network are publicly available at the following links.
\begin{itemize}
    \item \textbf{Our method:} \url{https://github.com/AntonioLonga/ETNgen}
    \item \textbf{STM:} \url{https://github.com/temporal-graphs/STM}
    \item \textbf{TAGgen:} \url{https://github.com/davidchouzdw/TagGen}
    \item \textbf{Dymond:} \url{https://github.com/zeno129/DYMOND}
\end{itemize}

\bibliography{biblio}
\bibliographystyle{unsrt}

\newpage

\appendix

\renewcommand\Authfont{\fontsize{12}{14.4}\selectfont}
\renewcommand\Affilfont{\fontsize{10}{10.8}\itshape}

\title{Supplementary Information\\
"Generating surrogate fine-grained temporal networks"
}
\date{\vspace{-5ex}}



\maketitle

\gdef\thefigure{\arabic{figure}}
\gdef\theequation{\arabic{equation}}
\gdef\thetable{\arabic{table}}
\setcounter{figure}{0}
\setcounter{equation}{0}
\setcounter{table}{0}

\renewcommand{\thefigure}{S\arabic{figure}}
\renewcommand{\thesection}{S\arabic{section}}
\renewcommand{\thetable}{S\arabic{table}}


\changed{\section{Topological and dynamical metrics details}}
In the main text we use \changed{seventeen} different topological metrics and three dynamical metrics to compare the original graphs with the synthetic ones. 
The topological metrics can be divided into \changed{thirteen time-dependent metrics,} which (except for contact durations) are computed for each temporal layer as it was a static network, and for which we report distributions over temporal layers (or over slots of one hour length, where indicated):
\begin{itemize}
\item \textbf{Density.} The ratio of edges in the graph versus the number of edges if it was a complete graph \cite{zeno2021dymond}.
\item \textbf{Interacting individuals.} The number of individuals that are interacting~\cite{starnini2012random}.
\item \textbf{New conversations.} The number of conversations starting at each specific timestamp~\cite{starnini2012random}.
\item \textbf{\changed{Hour} S-metric.} A measure of the extent to which a graph has a hub-like core, maximized when high-degree nodes are connected to other high-degree nodes~\cite{li2005towards}. \changed{ This is not computed on singular layers but on the networks resulting from aggregating slots corresponding to 1 hour.}
\item \textbf{Hour clustering coefficient.}  The ratio of the number of closed triplets to the total number of open and closed triplets \cite{luce1949method,wasserman1994social}. \changed{This is computed on the networks resulting from aggregating slots of 1 hour.}
\item \textbf{Hour assortativity.} The degree-degree correlation of nodes that are connected~\cite{newman2002assortative}. \changed{This is computed on the networks resulting from aggregating slots of 1 hour.}
\item \textbf{\changed{Hour} average shortest path length.} The average shortest path length for all possible pairs of nodes of the largest connected component for each 1-hour-aggregated network \cite{holme2012temporal}. \changed{This is computed on the networks resulting from aggregating slots of 1 hour.}
\item \changed{\textbf{Hour modularity.} First a partition of the 1-hour-aggregated network is obtained using the Louvain algorithm~\cite{blondel2008fast} and then the modularity is computed according to Clauset et al.~\cite{clauset2004finding}.}
\item \changed{\textbf{Hour betweenness centrality (weighted and unweighted).} Nodes centrality averaged over all nodes of the one-hour-long-aggregated networks~\cite{brandes2001faster}.}
\item \changed{\textbf{Hour closeness centrality.} Nodes centrality averaged over all nodes of the one-hour-long-aggregated networks~\cite{freeman2002centrality}. }
\item \changed{\textbf{Number of connected components.} Number of subnetworks each snapshot is divided into. }
\item \textbf{Duration of contacts.} The mean duration (in timestamps) of interactions between each couple of nodes~\cite{starnini2012random}.
\end{itemize}
And \changed{four time-aggregated metrics (computed over the aggregate networks)}:
\begin{itemize}
\item \changed{\textbf{Betweenness centrality (weighted and unweighted).} Centrality of each node in the aggregated~\cite{brandes2001faster}.}
\item \changed{\textbf{Closeness centrality.} Centrality of each node in the aggregated~\cite{freeman2002centrality}. }
\item \textbf{Edge strength.} Weight of each edge in the aggregated~\cite{starnini2012random}.
\end{itemize}

The dynamical metrics are obtained starting from two dynamical processes, a random walk and a spreading process. For random walk we use:
\begin{itemize}
    \item \textbf{Coverage.} The number of (distinct) visited nodes  starting from a random node at an initial timestamp \cite{starnini2012random}. The simulation is repeated 1000 times using a random initial node and the initial time is set equal to the first timestamp. 
    \item \textbf{Mean First Passage Time (MFPT).} The average time taken by the random walker to arrive for the first time at a specific node $i$, starting from a random initial position $j$ in the network~\cite{starnini2012random}. We consider each couple of nodes $(i,j)$ in the network and repeat the simulation five times for each of them.
\end{itemize}

The spreading process is a SIR model and we compute the following metric:
\begin{itemize}
    \item \textbf{Reproduction value $R_0$.} The average number of individuals infected by the first one, with a single random node infected as seed.
\end{itemize}


\section{Execution time comparison}
\label{sec:comp_time}

The egocentric perspective, that ignores interactions among neighbors of each ego node, implies a huge simplification with respect to mining standard motifs. Traditional techniques for motifs mining indeed rely on an isomorphism test for assessing sub-network equivalence, which is a major bottleneck for the entire procedure. For this reason, standard motifs mining techniques usually limit the search to small motifs containing a handful of nodes. 
The strength of \textit{ETN-gen} lies in the possibility of encoding neighborhoods into a unique bit vector, boiling down sub-network equivalence to bit vector matching. This hence results in a very computationally efficient model, and the time required for network generation is drastically lower than that of the other methods. This is evident from Table \ref{tab:time}, where we report the time (in seconds) required to generate networks for the three face-to-face datasets with all the methods. \textit{ETN-gen} is more than 15 times faster than the fastest state-of-the-art method on each network, and there is a difference in time of three orders of magnitude with the slowest one.


\begin{table}[h!]
\begin{center}
\begin{tabular}{r | c c c }
        & Hospital & Workplace & High School \\
        \hline
{\large \textcolor{col_e}{\textbf{\textit{ ETN-gen}}}} &   \textbf{$17s$}   & \textbf{$52s$}   &  \textbf{$22s$}  \\
{\large \textcolor{col_d}{\textbf{\textit{Dymond}}}}  & $10h$ & $23min$ &  $88h$ \\
{\large \textcolor{col_s}{\textbf{\textit{STM}}}}    & $23min$  & $16min$ &  $26min$ \\
{\large \textcolor{col_t}{\textbf{\textit{TagGen}}}} & $7h$ & $2h$ & $6h$ 
\end{tabular}
\caption{\textbf{Execution time.} Time (in seconds (s), minutes (min) or hours (h)) required to train and generate networks with each method on three different networks.}
\label{tab:time}
\end{center}
\end{table}

\subsection{Computational complexity and space complexity}\label{si:complex}

In this section, we report the time and space complexity of our model.\\

\textbf{Time complexity.} As depicted in Figure \ref{fig:fig0} of the Main Text, the method can be decomposed in four steps: 1) mine Egocentric Temporal Neighborhoods, 2) build a local probability distribution, 3) generate a provisional layer for each timestamp and 4) validate layer connections. Longa et al.\cite{longa2021efficient} proved that the computational cost to count all  Egocentric Temporal Neighborhoods in a graph is equal to $\mathcal{O}(n \cdot m_{orig} \cdot d^k \cdot \log d^k)$, where $n$ is the number of nodes, $m_{orig}$ is the number of timestamps in the original network, $d$ is the maximal degree of the network and $k$ is the length of the temporal neighborhood.  The second step can be done in linear time with respect to the size of the mined Egocentric Temporal Neighbors. In the third step, we query the local probabilistic model in constant time for each node for each timestamp of the generated network ($m_{gen}$), thus the complexity is $\mathcal{O}(n \cdot m_{gen})$. Finally, in the validation step, for each node and each timestamp we have to go through each edge (there are at most $d$ of them), with a complexity of $\mathcal{O}(n \cdot m_{gen} \cdot d)$. The overall complexity is thus $\mathcal{O}(n \cdot m_{orig} \cdot d^k \cdot \log d^k + n \cdot m_{gen} \cdot d)$. 
Note that for reasonable values of $k$, $d^k$ is independent of the size of the network, so that the overall complexity is $\mathcal{O}(n \cdot m_{gen})$, assuming that $m_{gen} \gg m_{orig}$.
\\

\textbf{Space complexity.} The space complexity of the method is dominated by the storage of local probabilistic models. Storing a single Egocentric Temporal Neighborhood signature of length $k$ costs $\mathcal{O}((k+1) \cdot d)$, where $d$ is the maximal degree of the network \changed{(it's worth noting that a relationship exists between the maximal degree and the number of edges)}. The number of Egocentric Temporal Neighborhoods is the number of all possible ordered sequences of $k-bit$ strings of length $d$, which corresponds to ${d + 2^k -1 \choose d}$ and is loosely upper bounded by $2^{kd}$. The overall space complexity is thus $\mathcal{O}(2^{(k+1)d})$.  As discussed in the case of time complexity, for reasonable values of $k$, $d$ is independent of the size of the network, so that the space complexity does not depend on it.

\section{Scalability}\label{si:scalability}

To show the scalability of our approach we extend the analysis to other seven networks, briefly described below.

\begin{itemize}
    \item \textbf{High school 2~\cite{10.1371/journal.pone.0107878}.} The dataset has been collected in 2012 in Lycée Thiers, Marseilles, France, over seven days (Monday to Tuesday of the following week). It contains interactions among students in five different high school classes.  Number of edges: $2220$, number of nodes: $180$. As stated by the research group responsible for the data collection, a signed informed consent was obtained for each study participant (all involved students were at least 18). Moreover, the study was approved by the “Commission Nationale de l’Informatique et des Libertés” (CNIL, http://www.cnil.fr), the French national body responsible for ethics and privacy, and by the high school authorities. More details can be found in the paper describing the data collection~\cite{10.1371/journal.pone.0107878}.
    \item \textbf{High school 3~\cite{mastrandrea2015contact}.} The dataset has been collected in 2013 in Lycée Thiers, Marseilles, France, over five days in December. It contains interactions among students in nine different high school classes. Number of edges: $5818$, number of nodes: $327$. As stated by the research group responsible for the data collection, a signed informed consent was obtained for each study participant (all involved students were at least 18). Moreover, the study was approved by the “Commission Nationale de l’Informatique et des Libertés” (CNIL, http://www.cnil.fr), the French national body responsible for ethics and privacy, and by the high school authorities. More details can be found in the paper describing the data collection~\cite{mastrandrea2015contact}.
    \item \textbf{Primary school~\cite{stehle2011high}.} The dataset has been collected in a primary school in France, over two days in October 2009. It contains interactions among 232 children and 10 teachers. Number of edges: $8317$, number of nodes: $242$. As stated by the research group responsible of the data collection, the “Commission Nationale de l’Informatique et des Libertés” (CNIL, http://www.cnil.fr) and the “Comité de Protection des personnes” (http://www.cppsudest2.com/) were notified of the study. The study was also approved by the relevant academic authorities of the primary school in which the study took place. Finally, parents, teachers, and the director of the school expressed a verbal informed consent. More details can be found in the paper describing the data collection~\cite{stehle2011high}. 
    \item \textbf{SMS 1~\cite{sapiezynski2019interaction}.} The dataset represents SMSs among  university freshmen students in the Copenhagen University. Number of edges: $697$, number of nodes: $568$. The dataset was collected within the Copenhagen Network Study and the data collection was approved by the Danish Data Supervision Authority. Each study participant was asked to sign an informed consent. 
    \item \textbf{SMS 2~\cite{aharony2011social}.} The dataset represents SMSs among members of a young-family residential living community adjacent to a major research university in North America. Number of edges: $153$, number of nodes: $85$. The dataset was collected within the Friends and Family Study and the data collection was approved by the Institutional Review Board (IRB). The participation was optional and each study participant was asked to explicitly adhere. 
    \item \textbf{Calls 1~\cite{sapiezynski2019interaction}.} The dataset represents phone calls among university freshmen students in the Copenhagen University. Number of edges: $605$, number of nodes: $525$. The dataset was collected within the Copenhagen Network Study and the data collection was approved by the Danish Data Supervision Authority. Each study participant was asked to sign an informed consent.
    \item \textbf{Calls 2~\cite{aharony2011social}.} The dataset represents phone calls among members of a young-family residential living community adjacent to a major research university in North America.
    Number of edges: $432$, number of nodes: $129$. The dataset was collected within the Friends and Family Study and the data collection was approved by the Institutional Review Board (IRB). The participation was optional and each study participant was asked to explicitly adhere.
\end{itemize}

Each face-to-face interaction network has been aggregated with a temporal resolution of five minutes, while SMS and phone calls networks have been aggregated within ten minutes. We opt for this different aggregations due to the natural sparsity of SMS and phone calls networks.

\begin{figure}[!h]
\begin{center}
\includegraphics[width=\textwidth]{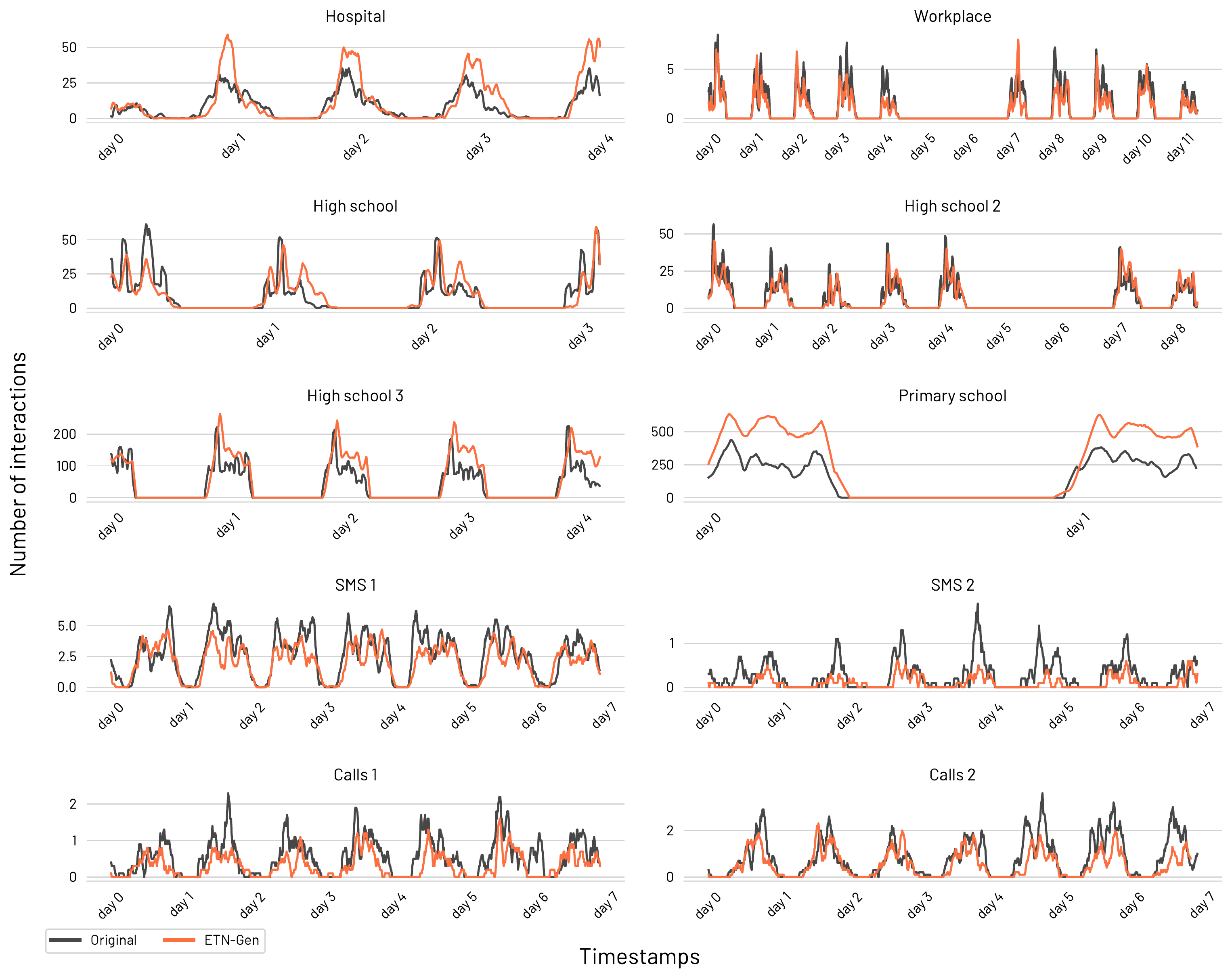}
\caption{\textbf{Number of interactions in the generated network for different datasets.} Each panel shows the number of interactions of the original (black curve) and \textit{ETN-gen} (orange curve) graphs. We use a temporal gap of 5 minutes for face-to-face interactions and 10 minutes for calls and SMS (intrinsically sparser networks). 
}
\label{fig:nb_inter_generalization_notag}
\end{center}
\end{figure}

In Figure \ref{fig:nb_inter_generalization_notag} we show the original number of interactions (in black) and those generated by our method (in orange) for each network. The figure clearly shows the ability of our method in mimicking day/night and week/weekend periodicity. Moreover, our algorithm perfectly operates with different network sizes in both number of individuals and temporal length. Finally, our method is
able to capture multiple picks within the same day, that could be associated to the period before and after lunch (i.e. high schools).


\section{\changed{Dynamical similarity over time}}
\label{sec:si:dyn}
\begin{figure}[!h]
     \centering
     \includegraphics[width=\textwidth]{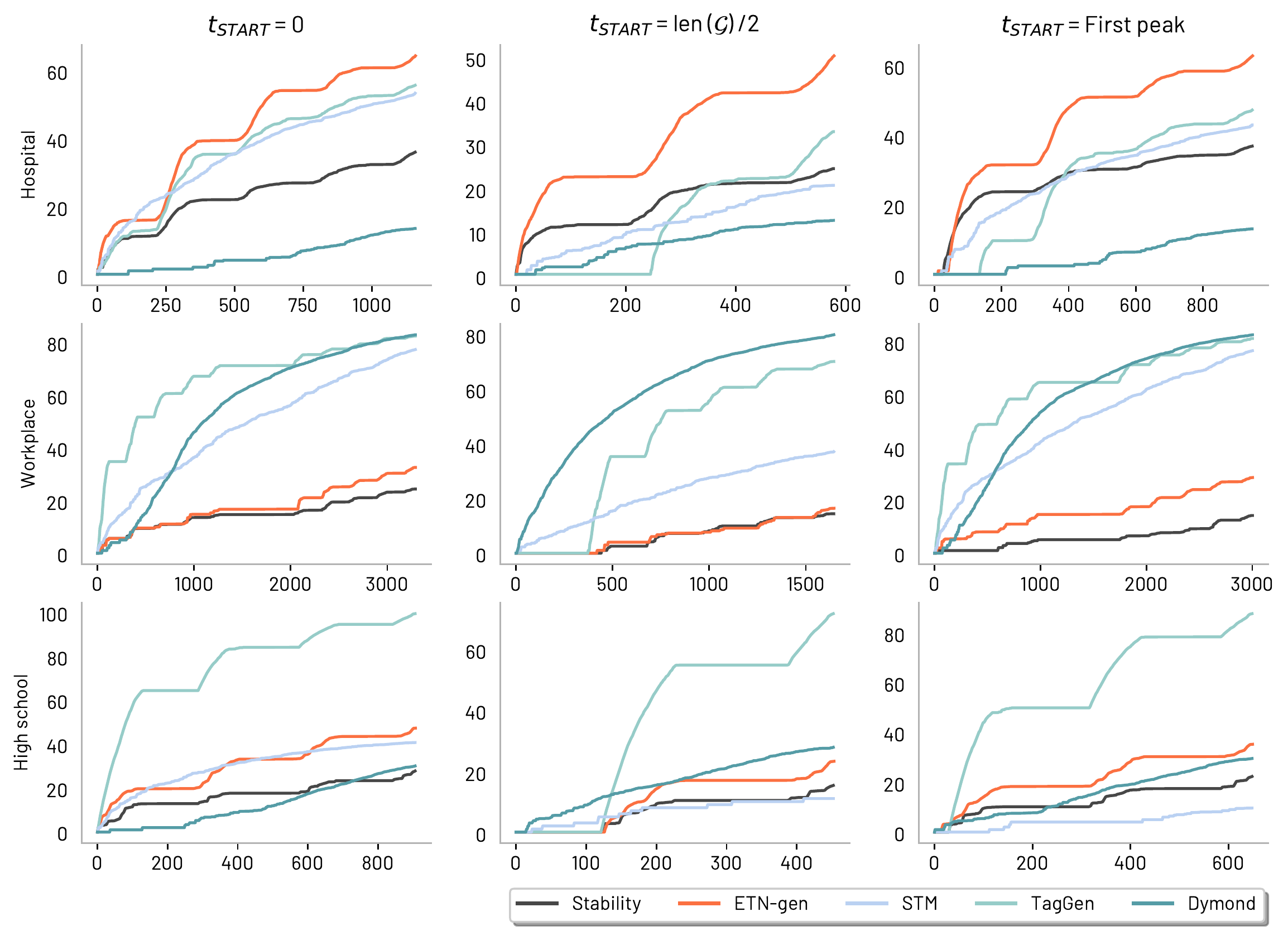}
    \caption{\changed{Random walk coverage in time in each generated network for three different starting points: time 0, T/2 and on the first peak (when the number of connections reaches a maximum).}}
    \label{fig:si:seen_nodes_RW}
\end{figure}

\begin{figure}[!h]
     \centering
     \includegraphics[width=\textwidth]{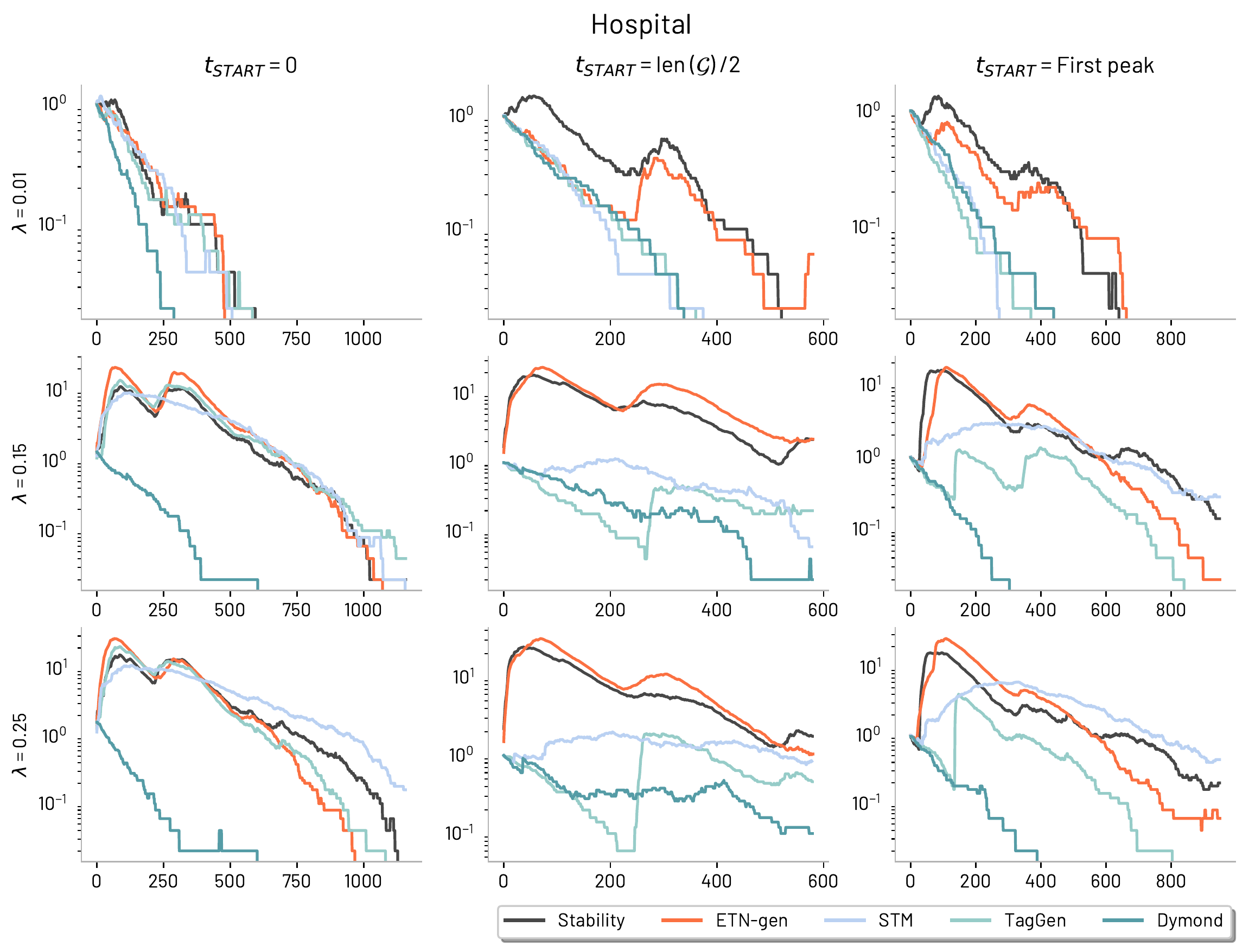}
    \caption{\changed{SIR number of infected nodes in time for the Hospital network for three different starting points: time 0, T/2 and on the first peak (when the number of connections reaches a maximum).}}
    \label{fig:si:nb_infected_lh}
\end{figure}

\begin{figure}[!h]
     \centering
     \includegraphics[width=\textwidth]{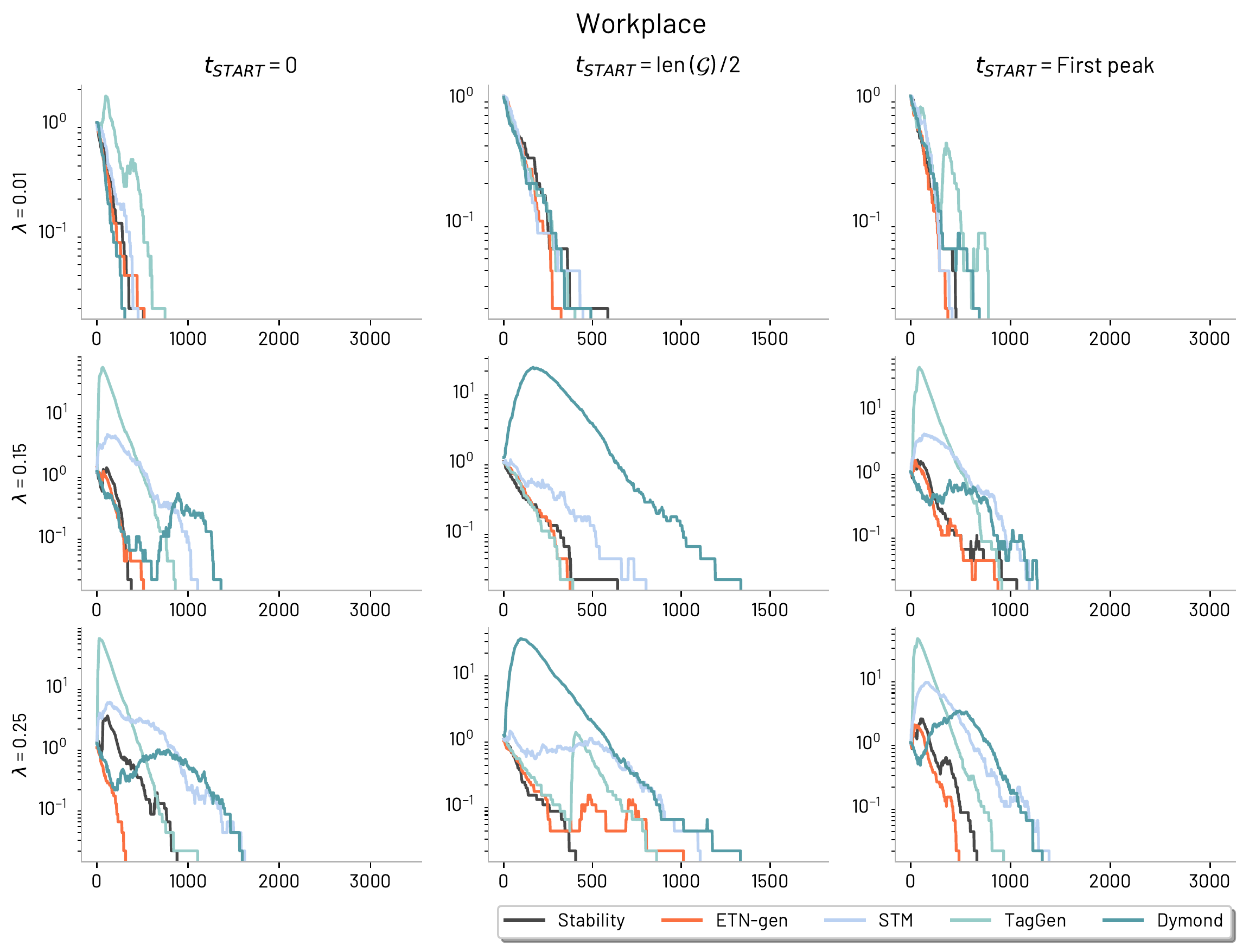}
    \caption{\changed{SIR number of infected nodes in time for the Workplace network for three different starting points: time 0, T/2 and on the first peak (when the number of connections reaches a maximum).}}
    \label{fig:si:nb_infected_invs}
\end{figure}

\begin{figure}[!h]
     \centering
     \includegraphics[width=\textwidth]{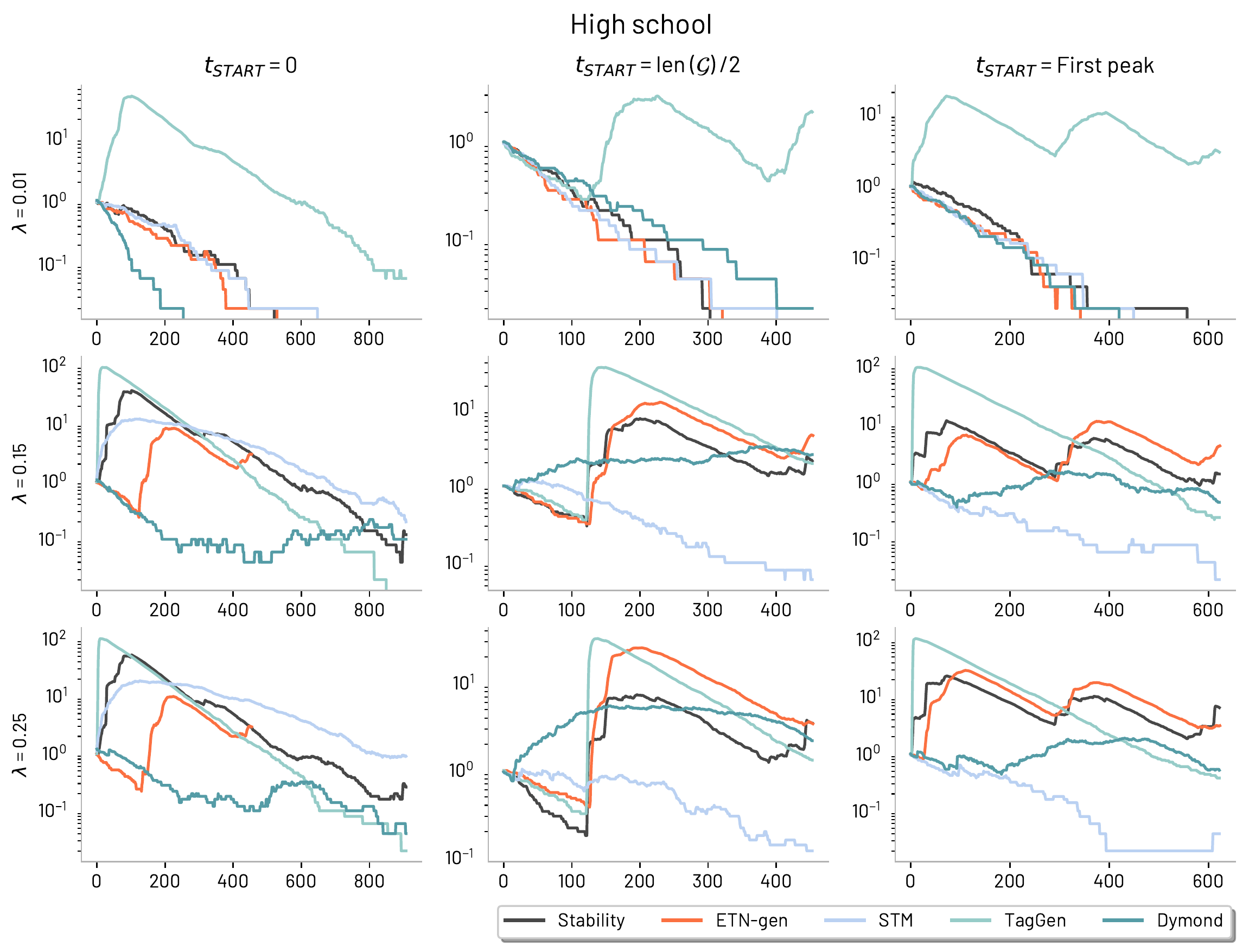}
    \caption{\changed{SIR number of infected nodes in time for the High school network for three different starting points: time 0, T/2 and on the first peak (when the number of connections reaches a maximum).}}
    \label{fig:si:nb_infected_hs}
\end{figure}

\changed{
For the same dynamical simulations shown in Section~\ref{sec:dynamic} of the Main Text we also show the evolution in time. For the random walk process we show how the number of newly visited nodes increases in time, see Figure~\ref{fig:si:seen_nodes_RW}. For the SIR spreading model we show how the number of infected changes in time, see Figures~\ref{fig:si:nb_infected_lh}, ~\ref{fig:si:nb_infected_invs}, and ~\ref{fig:si:nb_infected_hs}.
By eye inspection the processes simulated on \etngen networks show a temporal evolution very similar to those simulated on original graphs. If compared with the other generated networks, \etngen either shows the highest similarity or results comparable with them.
}

\clearpage


\section{Additional datasets: topological similarity}

In this section we show the effectiveness of \textit{ETN-gen} when using alternative temporal networks as original graphs. We test some additional face-to-face interactions networks, then we  consider remote communication interactions like SMS and calls networks, for a total of 10 different temporal datasets. 
We compare \textit{ETN-gen} results with \textit{TagGen} as a sole \changed{alternative method.} This choice is mainly due to time constraints: the other \changed{methods} would require a very long time to complete the generation process when dealing with these large datasets (see Section~\ref{sec:comp_time}). Moreover, \textit{TagGen} is the only state-of-the-art algorithm able to always reproduce the same exact number of nodes of the input network, and the only \changed{method} able to capture intrinsic periodicity of the network.

\subsection{Other face-to-face interactions networks}
\begin{figure}[!h]
\begin{center}
     \includegraphics[width=\textwidth]{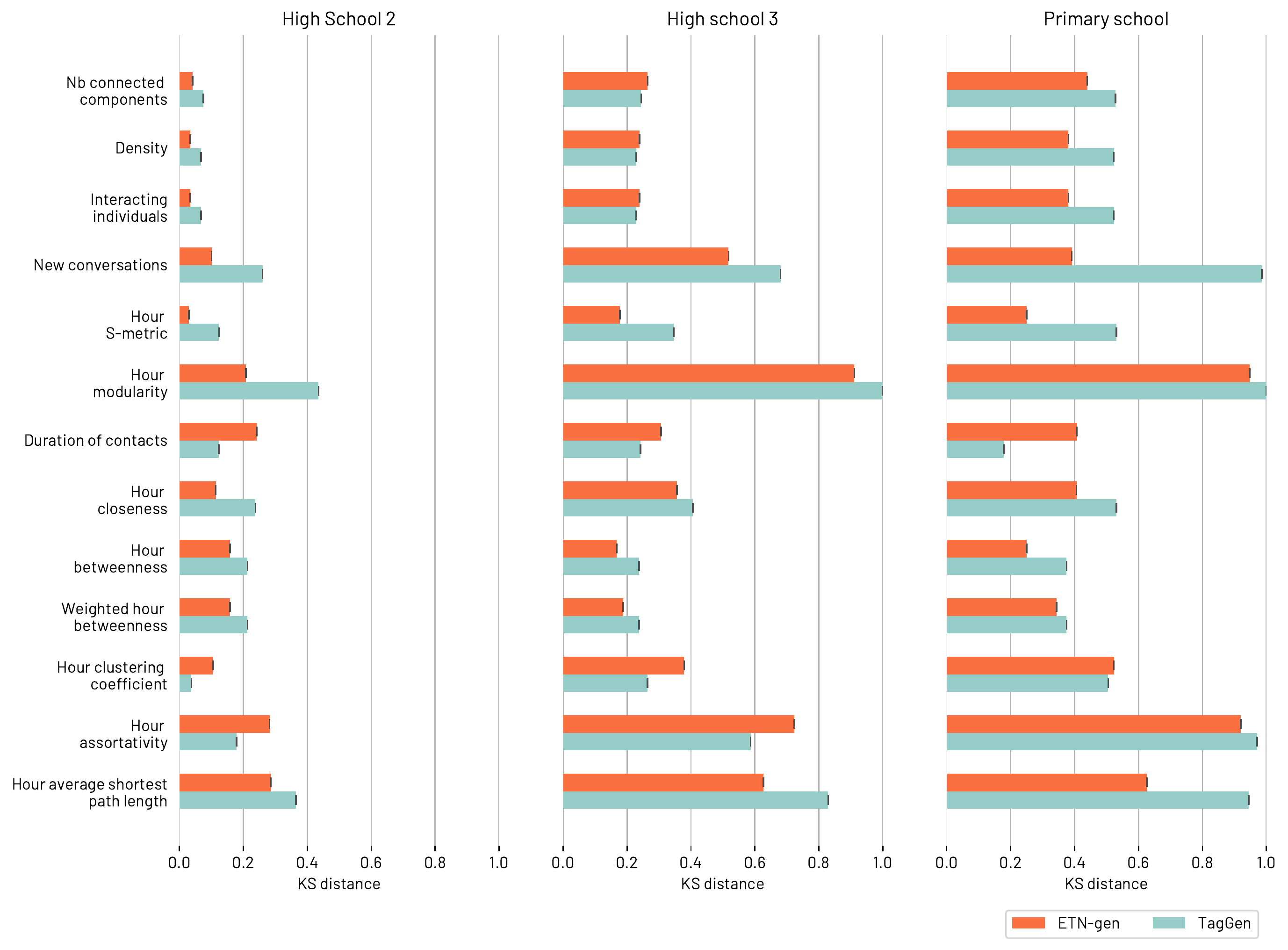}
    \caption{\changed{Kolmogorov-Smirnov distances applied to 10 different topological metric distributions of face-to-face interaction networks.}}
\label{fig:si:ks_sta_face_to_face}
\end{center}
\end{figure}

\begin{figure}[!h]
     \centering
     \includegraphics[width=\textwidth]{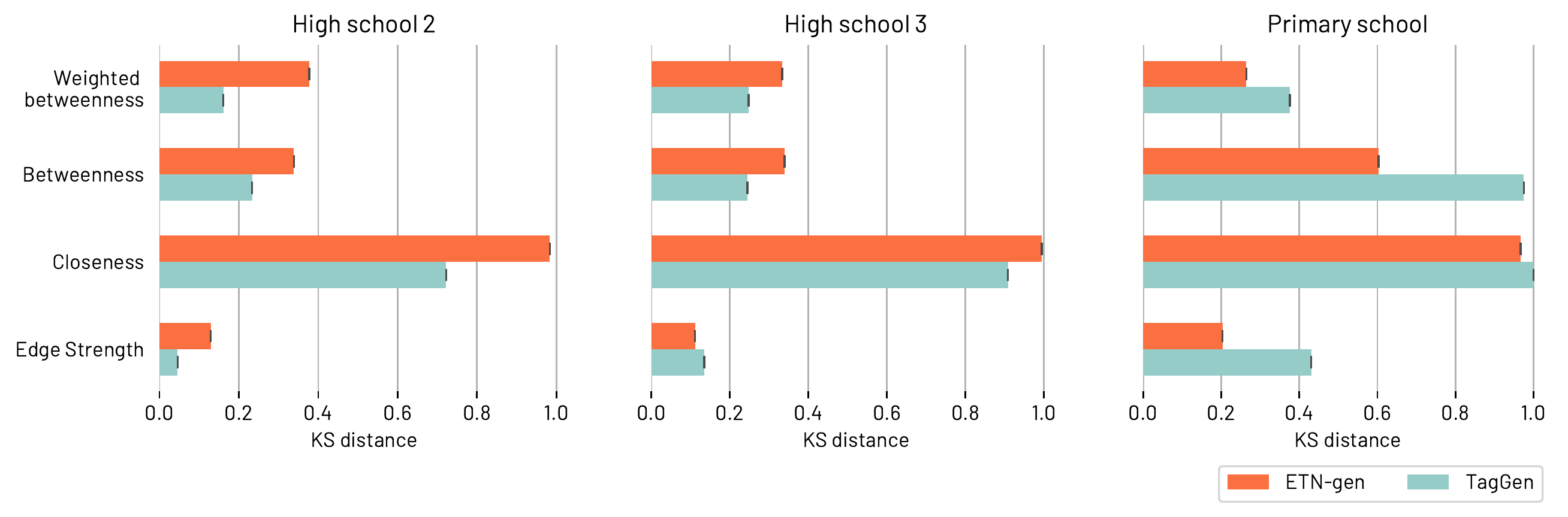}
    \caption{\changed{Topological similarity according to time-aggregated measures. Analogous to Figure \ref{fig:si:ks_sta_face_to_face} for measures on the aggregated networks.}}
    \label{fig:si:ks_sta_agg_face}
\end{figure}

In the Main Text we show how to generate surrogate networks that mimic three face-to-face interaction datasets, by making use of \textit{ETN-gen} and three other alternative methods. Here, we focus on three additional face-to-face interaction datasets, namely High school 2, High school 3 and Primary school. Kolmogorov-Smirnov distances of the metrics described in the Main Text (see \textit{Methods}) are shown in Figure \ref{fig:si:ks_sta_face_to_face}. 
\changed{In Figure \ref{fig:si:ks_sta_agg_face} we report the topological similarity between the aggregated original and generated networks.}

\changed{Analogously to the results of the Main Text, \etngen networks show a higher similarity to the original networks according to the time-dependent measures and a lower similarity according to the aggregated quantities.}

\clearpage
\subsection{SMS and phone calls networks}
\begin{figure}[!h]
     \centering
     \includegraphics[width=\textwidth]{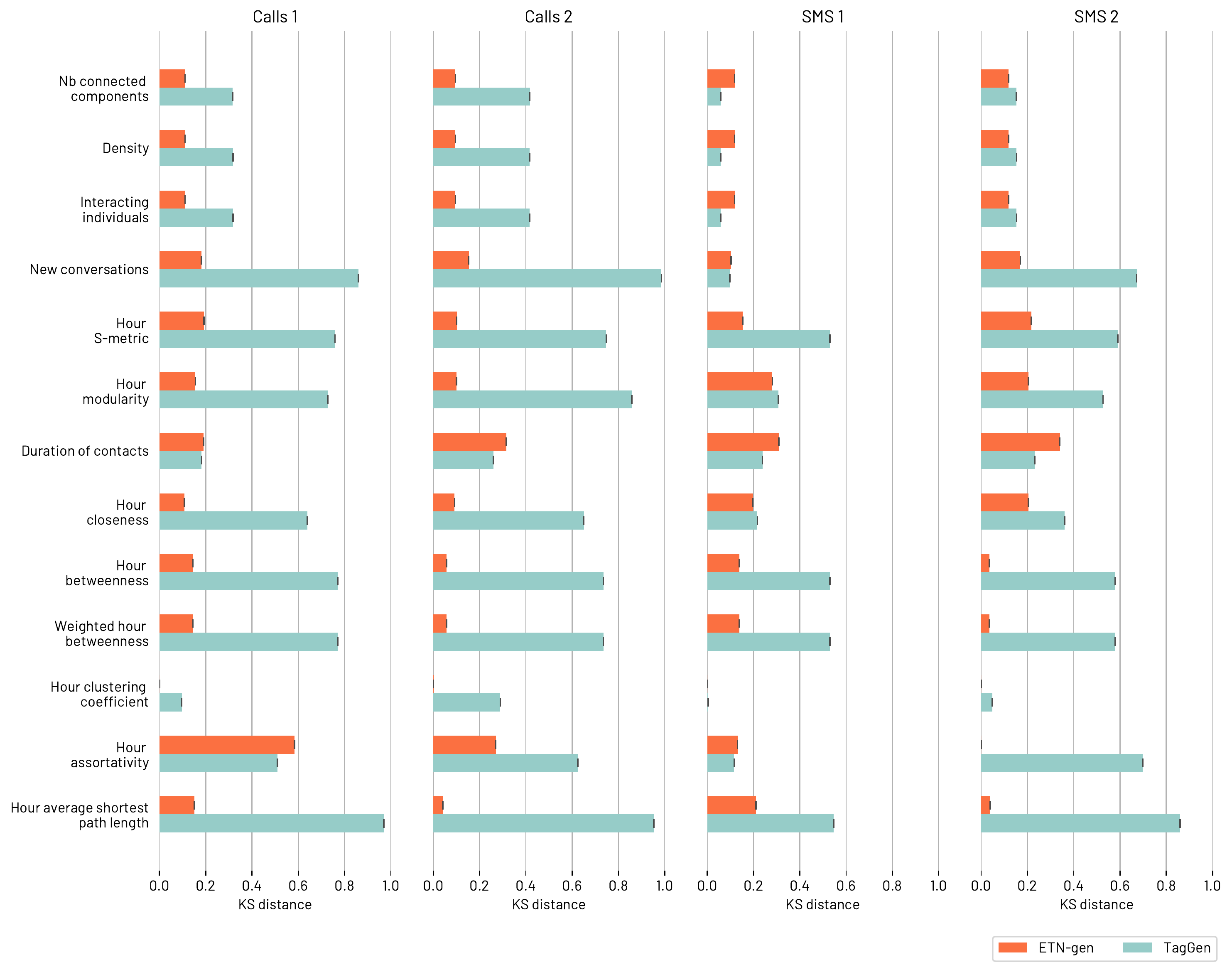}
    \caption{\changed{Kolmogorov-Smirnov distances applied to 10 different topological metric distributions of SMS and phone calls networks.}}
    \label{fig:si:ks_sta_sms_calls}
\end{figure}

\begin{figure}[!h]
     \centering
     \includegraphics[width=\textwidth]{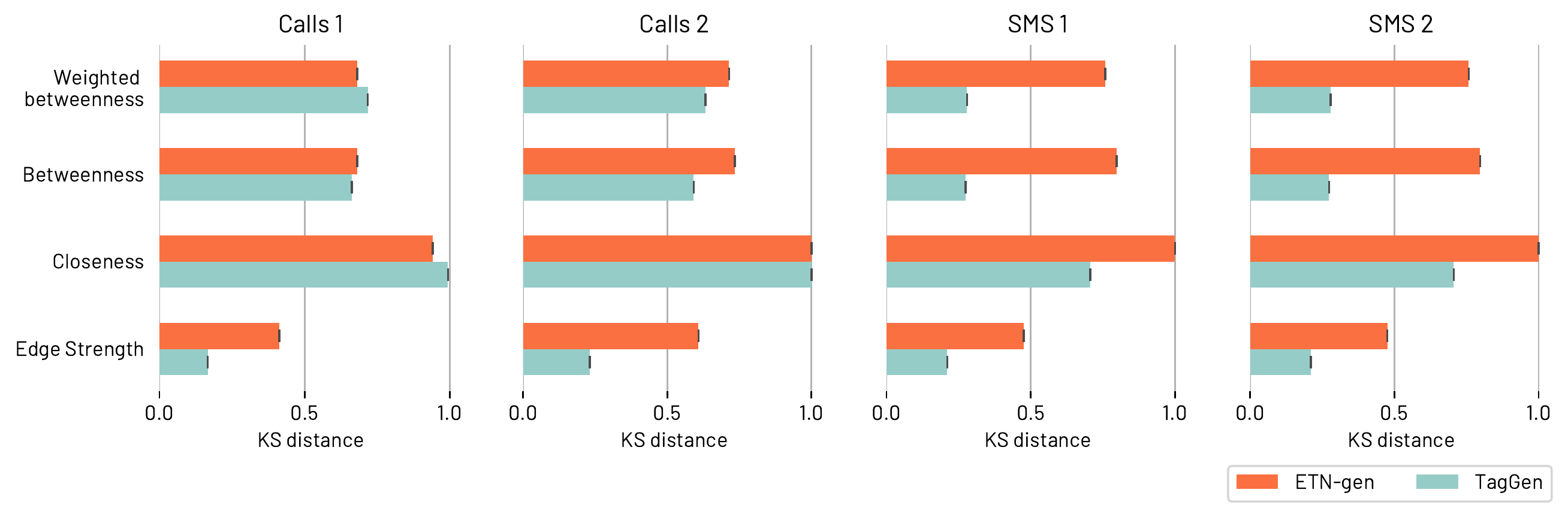}
    \caption{\changed{Topological similarity according to time-aggregated measures. Analogous to Figure \ref{fig:si:ks_sta_sms_calls} for measures on the aggregated networks.}}
    \label{fig:si:ks_sta_agg_sms}
\end{figure}

Figure \ref{fig:si:ks_sta_sms_calls} shows the distances among original and generated distributions of the chosen topological metrics. As expected, the methodology we propose is not able to capture metrics correlated to long-term memory. \changed{In Figure \ref{fig:si:ks_sta_agg_sms} we report the topological similarity between the aggregated original and generated networks.}

\changed{Analogously to the results of the Main Text, \etngen networks show a higher similarity to the original networks according to the time-dependent measures and a lower similarity according to the aggregated quantities.}

\clearpage

\section{Additional datasets: dynamical similarity}
Here we report Kolmogorov-Smirnov distances between original and generated networks in terms of random walks (coverage and mean first passage time) and the $R_0$ over a SIR model. 
\changed{We consider three different starting points: $t=0$, $t=len(G)/2$ (in the middle point of the temporal extension of the graph), and the time corresponding to the first peak of  connections (when the number of connections reaches a maximum). }

\subsection{Other face-to-face interaction networks}
\begin{figure}[!h]
     \centering
     \includegraphics[width=\textwidth]{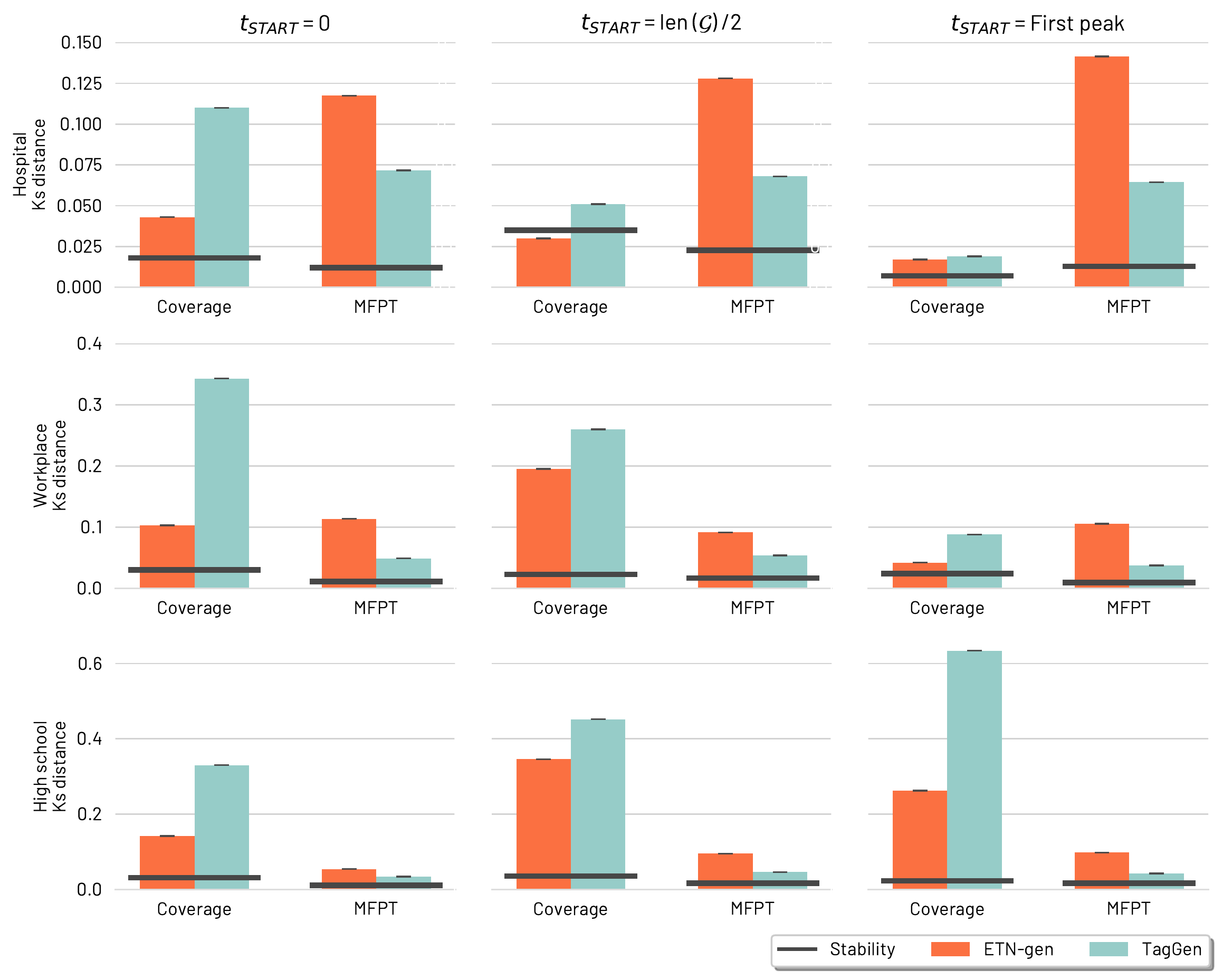}
    \caption{\changed{\textbf{Dynamical similarity in face-to-face networks: Random walk.} Kolmogorov-Smirnov distance between original and generated distributions of $R_0$ values on a SIR model simulation in face-to-face generated network for three different starting points: time 0, T/2 and on the first peak. Our method is represented in orange, while the solid black line shows the stability (i.e. the same simulation on the
original network).}}
    \label{fig:si:ks_sta_din_face_RW}
\end{figure}

\begin{figure}[!h]
     \centering
     \includegraphics[width=\textwidth]{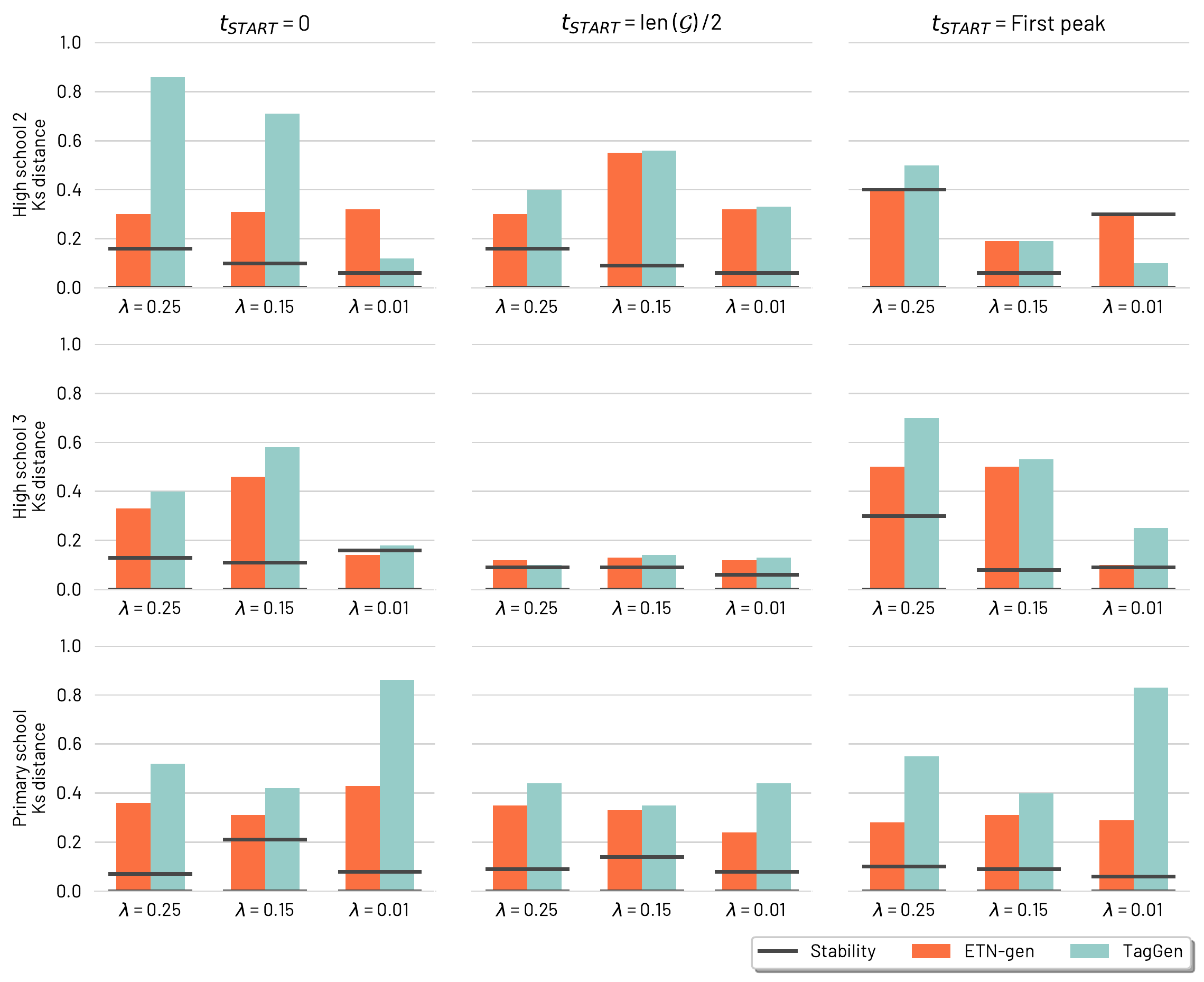}
    \caption{\changed{\textbf{Dynamic similarity in face-to-face networks: Spreading model.} Kolmogorov-Smirnov distance between original and generated distributions of $R_0$ values on a SIR model simulation in SMS and phone call generated networks for three different starting points: time 0, T/2 and on the first peak. Our method is represented in orange, while the solid black line shows the stability (i.e. the same simulation on the
original network).}}
    \label{fig:si:ks_sta_din_face_SIR}
\end{figure}

\changed{Figure \ref{fig:si:ks_sta_din_face_RW} shows that in the random walk process \etngen networks are able to capture  coverage  better than the \taggen ones, while these appear more similar to the original ones for what concerns the mean first passage time. There are no particular differences when changing the starting point.
Figure~\ref{fig:si:ks_sta_din_face_SIR} shows that for the spreading model the differences in the obtained distributions of $R_0$ between \etngen and \taggen are leveled out (and in some cases \etngen is performing better).
}

\clearpage
\subsection{SMS and phone calls networks}
\begin{figure}[!h]
     \centering
     \includegraphics[width=\textwidth]{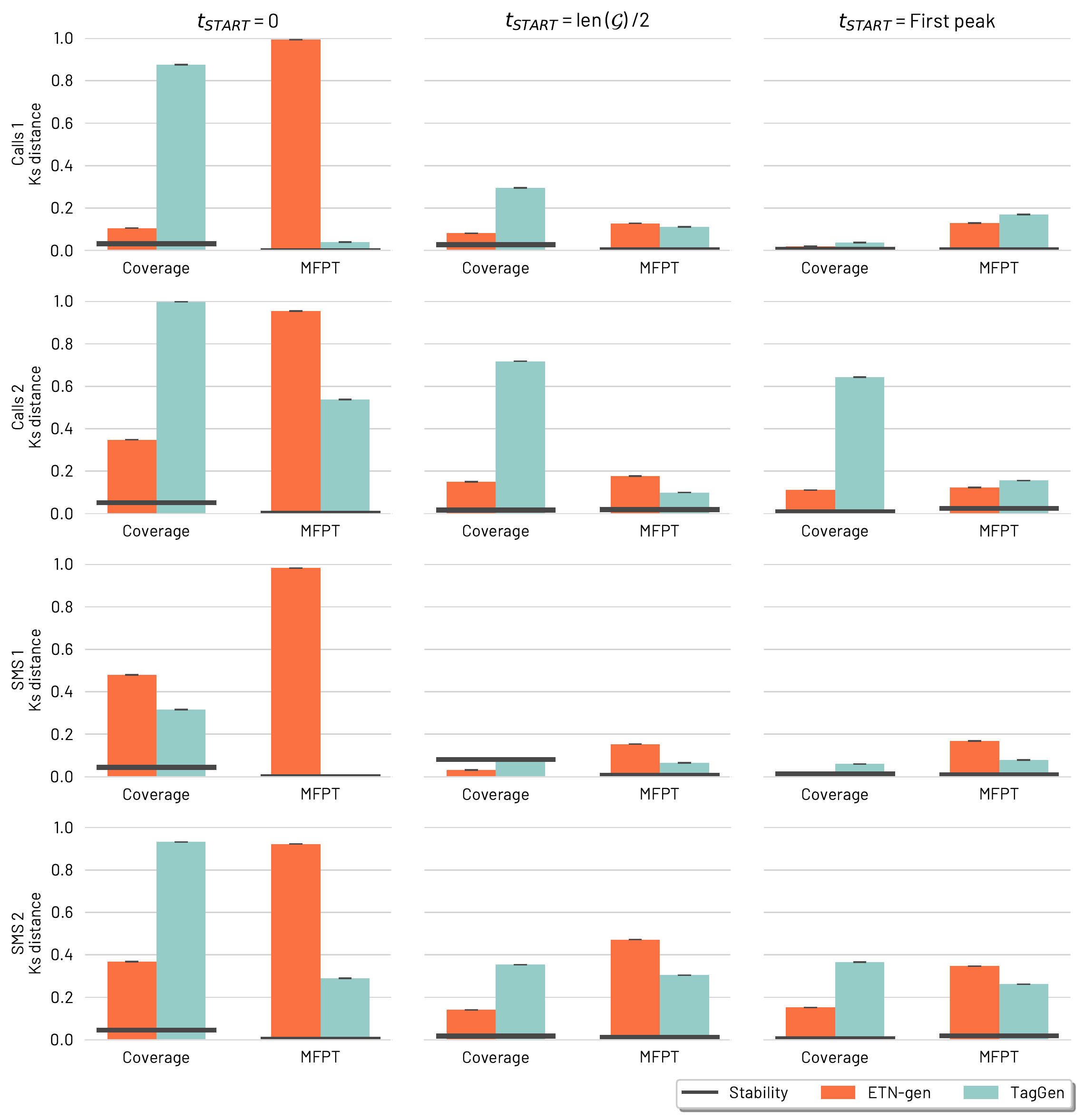}
    \caption{\changed{\textbf{Dynamic similarity in remote networks: Random walk.} Kolmogorov-Smirnov distance between original and generated distributions of coverage and mean first passage time in the random walk model in face-to-face generated networks for three different starting points: time 0, T/2 and on the first peak. Our method is represented in orange, while the solid black line shows the stability (i.e. the same simulation on the original network).}}
    \label{fig:si:ks_sta_din_sms_calls_RW}
\end{figure}

\begin{figure}[!h]
     \centering
     \includegraphics[width=\textwidth]{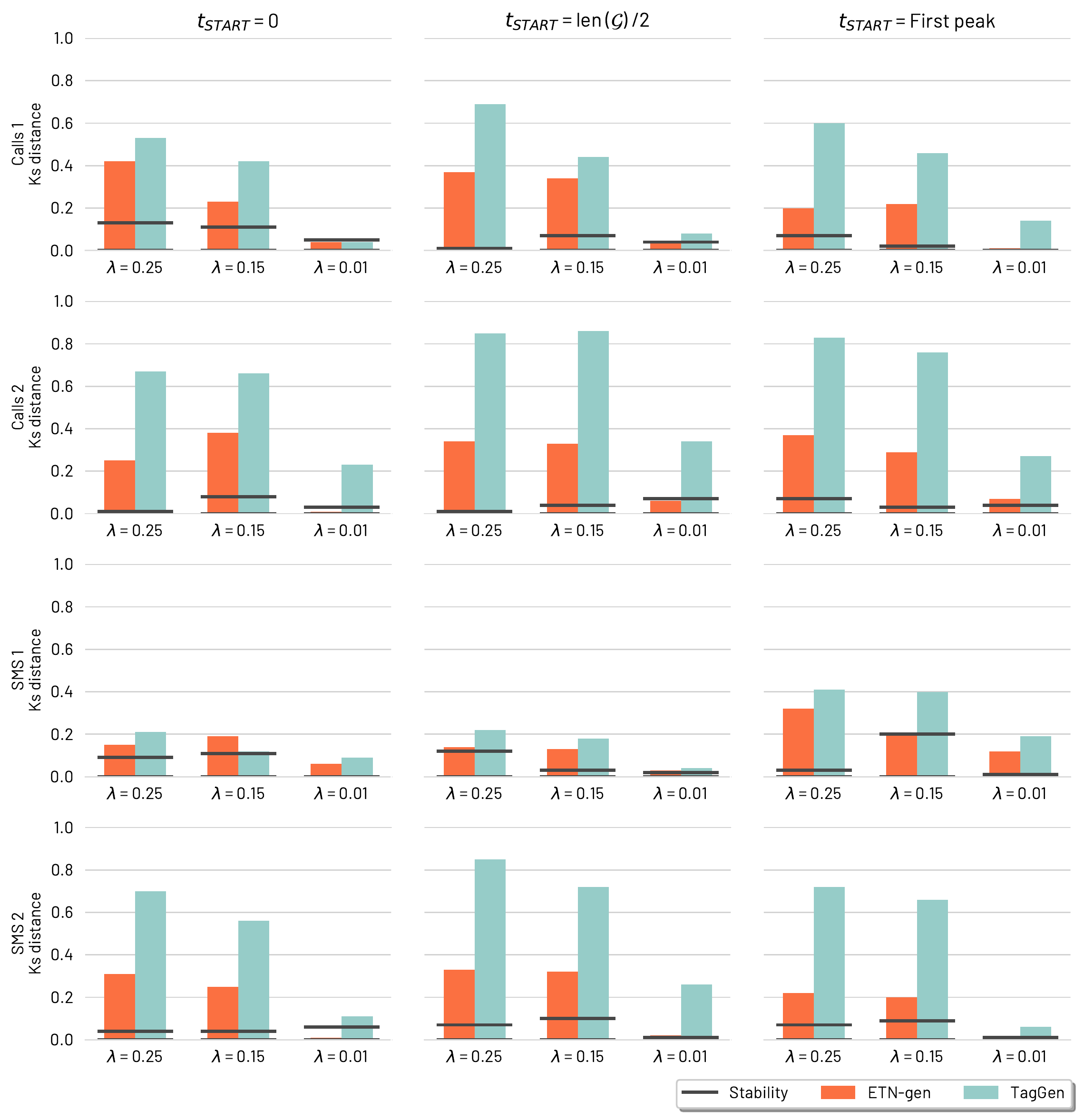}
    \caption{\changed{\textbf{Dynamic similarity in remote networks: Spreading model.} Kolmogorov-Smirnov distance between original and generated distributions of coverage and mean first passage time in the random walk model in  SMS and phone call generated networks for three different starting points: time 0, T/2 and on the first peak. Our method is represented in orange, while the solid black line shows the stability (i.e. the same simulation on the original network).}}
    \label{fig:si:ks_sta_din_sms_calls_SIR}
\end{figure}

\changed{The situation is a bit different for what concerns the remote interactions, which are characterized by more sparse networks. In particular from Figure \ref{fig:si:ks_sta_din_sms_calls_RW} we observe a lower similarity when the starting point is 0 and a much larger similarity for the other starting points. This is probably due to the noise inserted in the process when starting from one of the less connected temporal layers (time 0) in such sparse networks.
Figure~\ref{fig:si:ks_sta_din_sms_calls_SIR} shows a general good ability of \etngen networks to reproduce the $R_0$ distributions of the SIR model.
}

\newpage

\clearpage

\clearpage

\section{Topological similarity on one larger network}\label{si:bigdataset}

The purpose of our algorithm is to capture a high temporal resolution, on the other hand, the state-of-the-art methods have been developed for daily (or weekly) snapshots on bigger networks. For this reason, we evaluate the performance of our algorithm in generating a temporal network with a lower temporal resolution (one day). In particular, we used the same metrics used in \cite{zeno2021dymond} on the DNC dataset\cite{pickhardt2018extracting}. The DNC dataset is the network of emails of the Democratic National Committee that was leaked in 2016. The network is composed of $1 579$ nodes, $33 378$ temporal edges and $3 911$ unique edges.

To evaluate the generated network, we used the same metrics used in \cite{zeno2021dymond}, in particular, they studied the similarity between the generated and original networks computing the Kolmogorov-Smirnov distance of two distributions for some metrics. They divided the metrics into two categories \textit{Graph Structure Metrics} (density of the snapshots, global and local clustering coefficient,  S-metric and  average shortest path length) and \textit{Node Behaviour Metrics} (temporal degrees, local clustering,  closeness centrality,  connected component size and  activity rate). For the \textit{Node Behaviour Metrics}, they consider the interquartile range (IQR) of values over time.

\begin{figure}[!h]
     \centering
     \includegraphics[width=\textwidth]{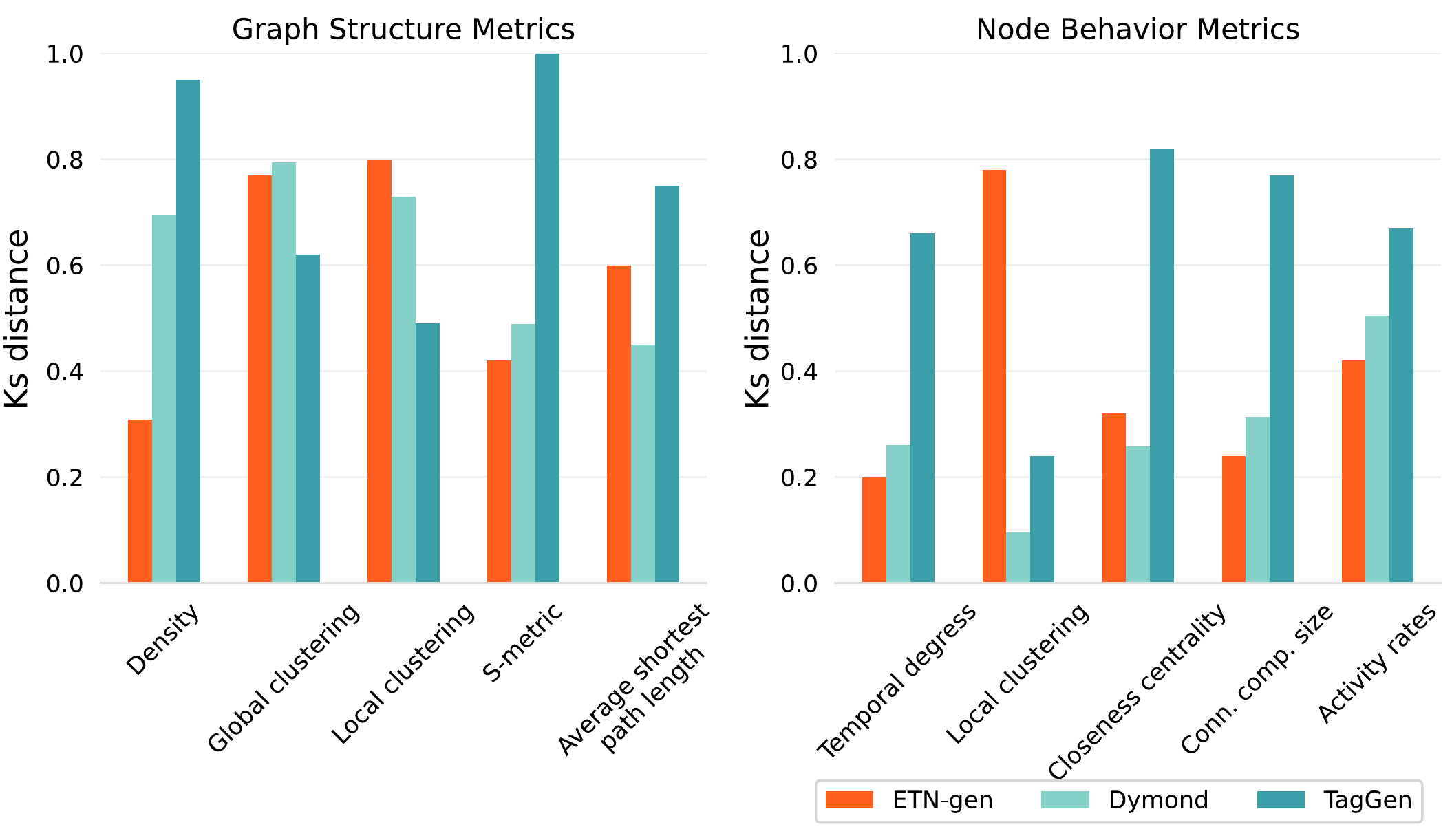}
    \caption{Comparison with the results obtained in \cite{zeno2021dymond}.}
    \label{fig:si:dymondres}
\end{figure}

Figure \ref{fig:si:dymondres} shows the Kolmogorov-Smirnov distances between generated networks and the original one. As depicted in the previous experiment, our algorithm is not capturing the global and local clustering coefficient distribution. However, we obtain the smallest distance with the original network in five out of ten metrics. More importantly, \etngen runs in the order of  minutes on a laptop, while the other methods are extremely expensive in terms of time (\dymond) or space (\taggen) requirements and cannot be easily run on consumable hardware.

\clearpage

\section{Varying K}\label{si:varing K}
\begin{figure}[!h]
     \centering
     \includegraphics[width=\textwidth]{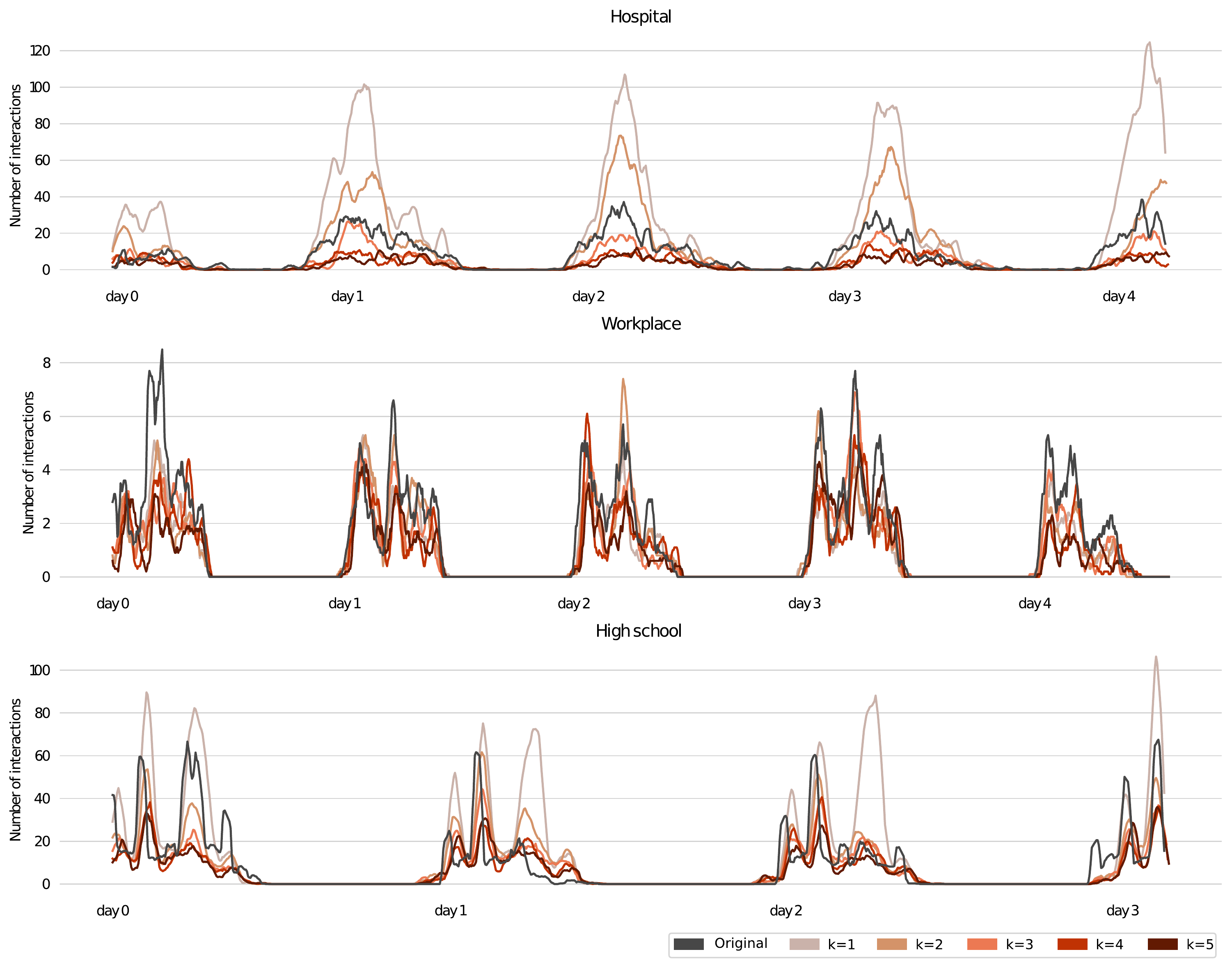}
    \caption{\changed{Number of interactions at varing $k$.}}
    \label{fig:si:varing_K_nbinter}
\end{figure}

\begin{figure}[!h]
     \centering
     \includegraphics[width=\textwidth]{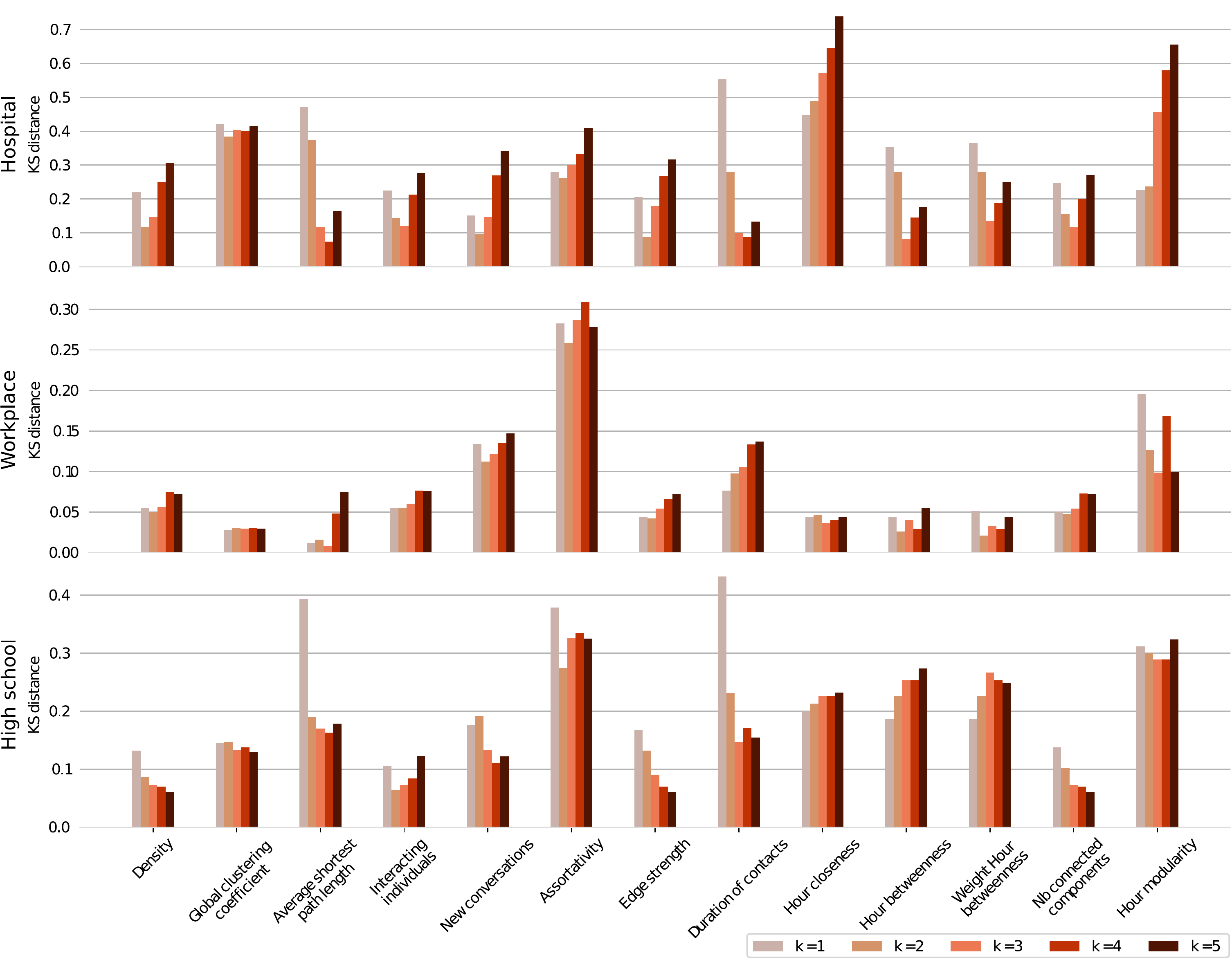}
    \caption{\changed{Topological similarity at varing $k$.}}
    \label{fig:si:varing_K_topo}
\end{figure}

\changed{
In this section we evaluate the performance of our method when $k$ varies between 1 and 5. 
Figure \ref{fig:si:varing_K_nbinter} shows the number of interactions and we observe that as much as $k$ increases, the average number of interactions tend to decrease. 
Figure \ref{fig:si:varing_K_topo} instead shows the topological similarity of several metrics with the original network. 
In general we observe that the differences in the topological measures are not significant in the range of k between 2 and 4 (see for instance density and number of connected components in all datasets), except for some individual cases (like hour closeness in the hospital dataset where clearly increasing $k$ decreases similarity, or the average shortest path length in the High school where the opposite is true).
}
As one may expect as far as $k$ increases, the execution time increases (see Table \ref{tab:exTime_varyingK}).\\

\begin{table}[h!]
    \centering
    \begin{tabular}{c|c c c c c}
        \textbf{k} & 1 & 2 & 3 & 4 & 5 \\
        \hline
        Hospital   & 8  & 17 & 25 & 40 & 55 \\
        Workplace  & 27 & 52 & 114 & 154 & 191 \\
        Hospital   & 10 & 22 & 42 & 59 & 85 \\
    \end{tabular}
    \caption{\changed{Execution time (in seconds) when varying K}}
    \label{tab:exTime_varyingK}
\end{table}

\clearpage

\section{Multiple versus single probabilistic model}\label{si:multiple_dict}

\begin{figure}[!h]
     \centering
     \includegraphics[width=\textwidth]{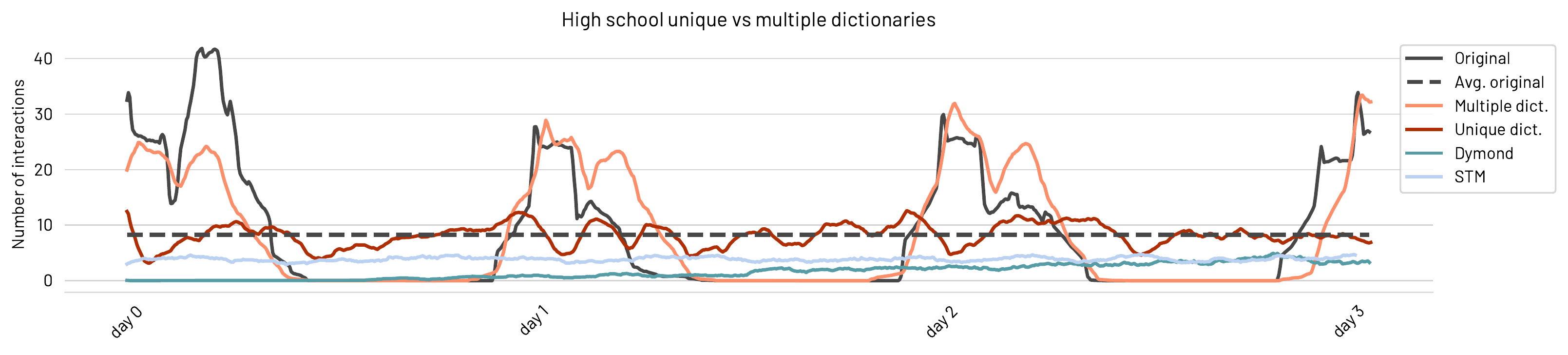}
    \caption{Number of interactions on high school network generated by multiple and unique local probabilistic models in \etngen and by the other methods.}
    \label{fig:si:multi_dict}
\end{figure}

In this section, we show that using a unique probabilistic model does not capture the daily/night periodicity. However, we are able to capture the average number of interactions.\\
In the first panel of Figure \ref{fig:si:multi_dict}, it is shown the number of interactions of the original network (in black), the one generated by our method using multiple local probabilistic models (in orange), the number of interactions of the generated network with a unique local probabilistic model (in red), and those generated by \dymond and \stm. It is worth mentioning that \etngen with a unique probabilistic model matches the average number of interactions, while the other methods do not. In particular, the original average number of interactions is $8.27$, while, our method has an average number of interactions equal to $8.27$ and $8.19$ for multiple and unique probabilistic models, respectively. On the other hand, the average number of interactions of \dymond and \stm are $1.65$ and $3.89$. \\
In conclusion, it is true that using multiple probabilistic models stores more information of the input network. However, even using a unique probabilistic model, \etngen performs better than the other state-of-the-art models.

\clearpage
\section{\changed{Results on number of interactions in time, number of nodes, and topological similarities using k equal to 3}}\label{si:k3}
\begin{figure}[!h]
     \centering
     \includegraphics[width=\textwidth]{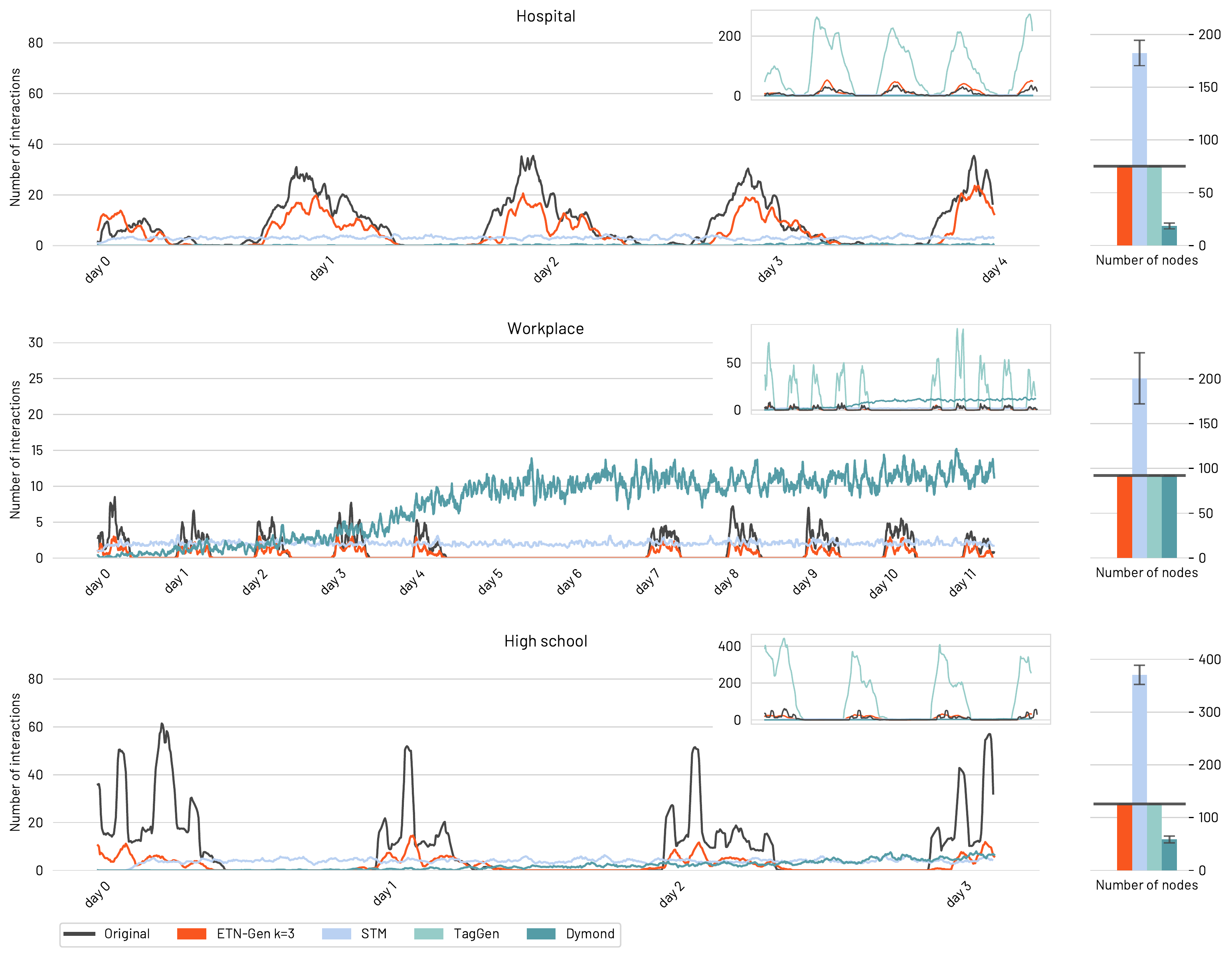}
    \caption{\changed{ \textbf{Number of interactions in time and number of nodes in the original and surrogate networks.} Each color represents a different generation algorithm, while the original graph is depicted in black. The insets depict the same curves with a different y-scale for visibility (the results obtained for TagGen are only reported there). The orange line refers to \etngen with k equal to 3.}}
    \label{fig:si:nbinteractionk3}
\end{figure}

\begin{figure}[!h]
     \centering
     \includegraphics[width=\textwidth]{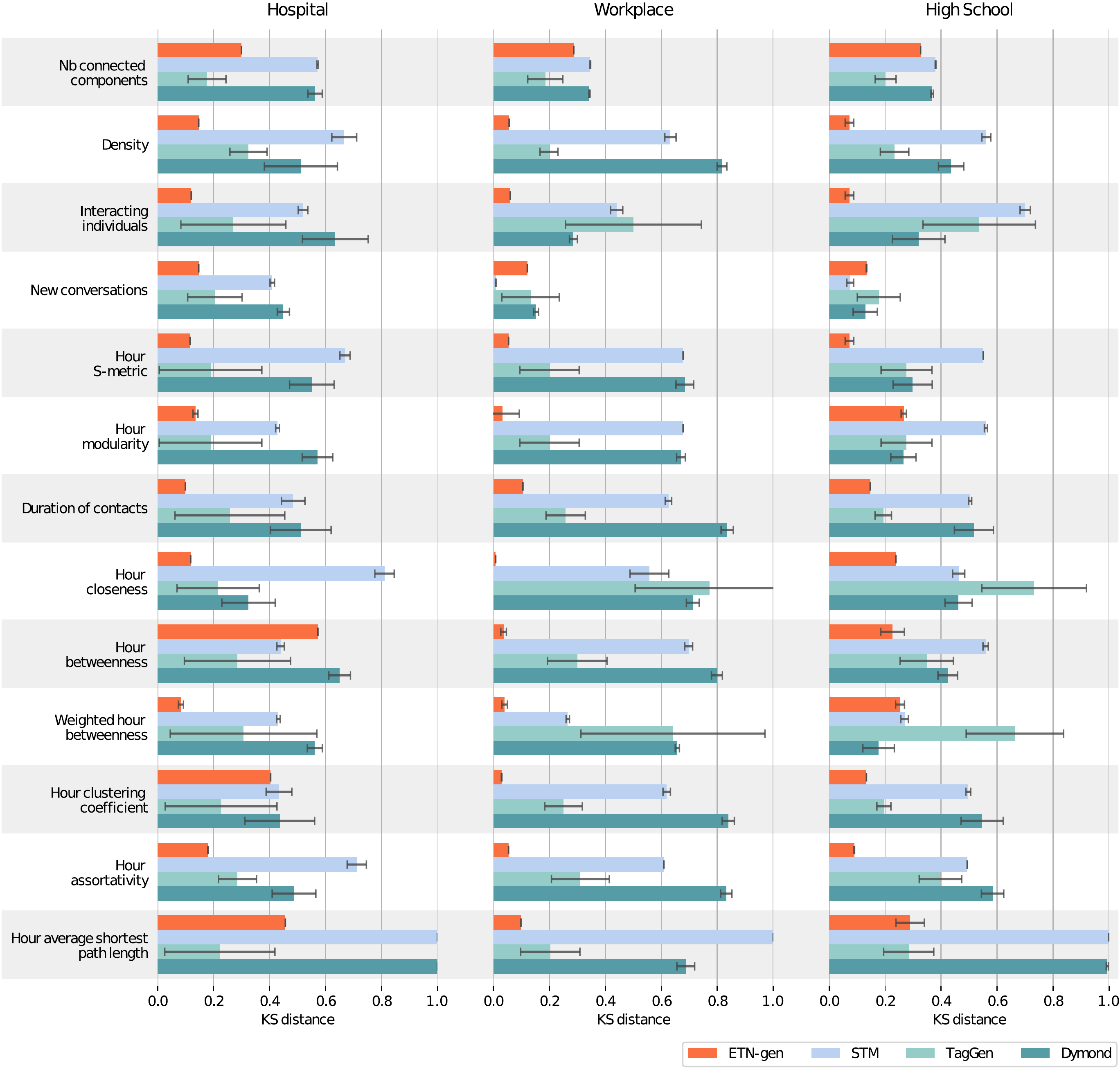}
    \caption{\changed{\textbf{Topological similarity according to time-dependent measures.} Similarity of the original network with those generated by \etngen (k equal to 3), \stm, \taggen and \dymond. Each bar reports the Kolmogorov-Smirnov distance between the two distributions (original and generated) for a specific structural metric. The shorter is a bar the more similar are the distributions. Standard deviations are obtained over 10 stochastic realizations of each network.}} 
    \label{fig:si:topologyk3}
\end{figure}

\begin{figure}[!ht]
\begin{center}
\includegraphics[width=\textwidth]{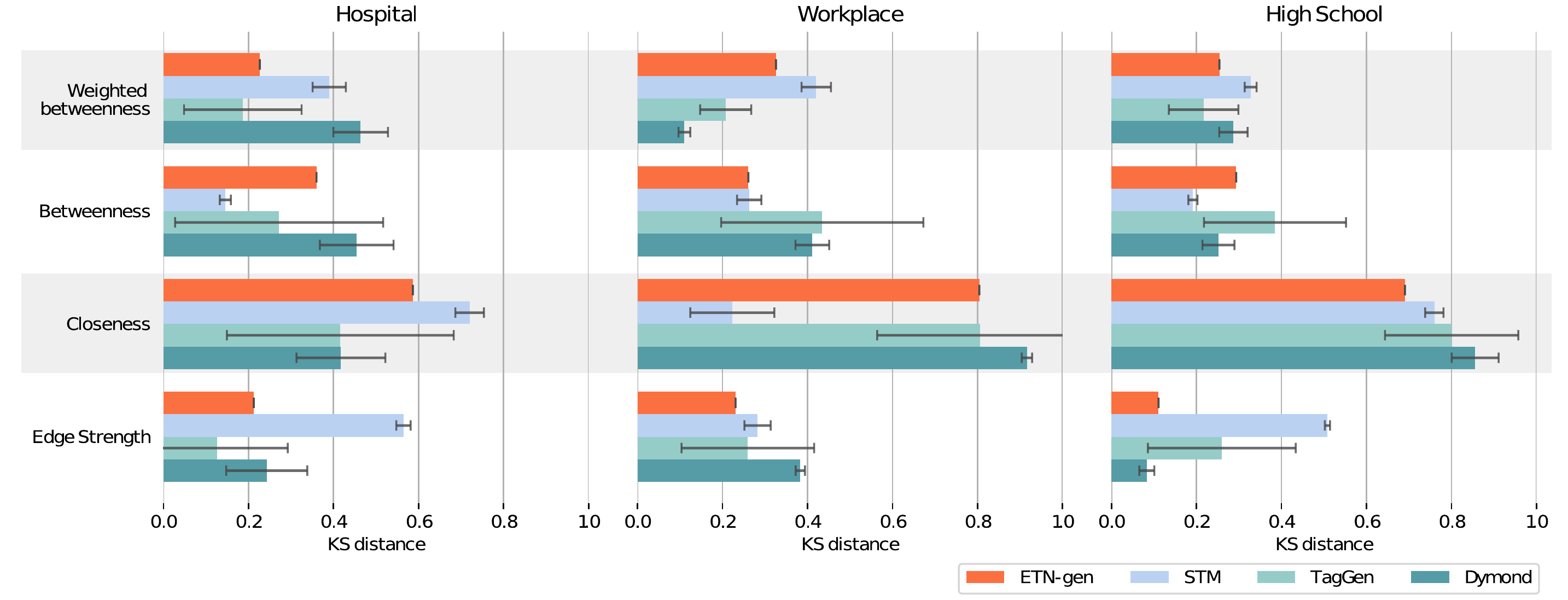}
\caption{\changed{\textbf{Topological similarity according to time-aggregated measures (\etngen with K = 3).} Analogous to Figure~\ref{fig:si:topologyk3} for measures on the aggregated networks.}}
\label{fig:top_kstest_timeaggregated3}
\end{center}
\end{figure}

\section{\changed{Mesoscale structures}}
\label{sec:si:mesoscale}

\subsection{\changed{Communities}}
\label{sec:si:communites}
\begin{figure}[!h]
     \centering
     \includegraphics[width=0.5\textwidth]{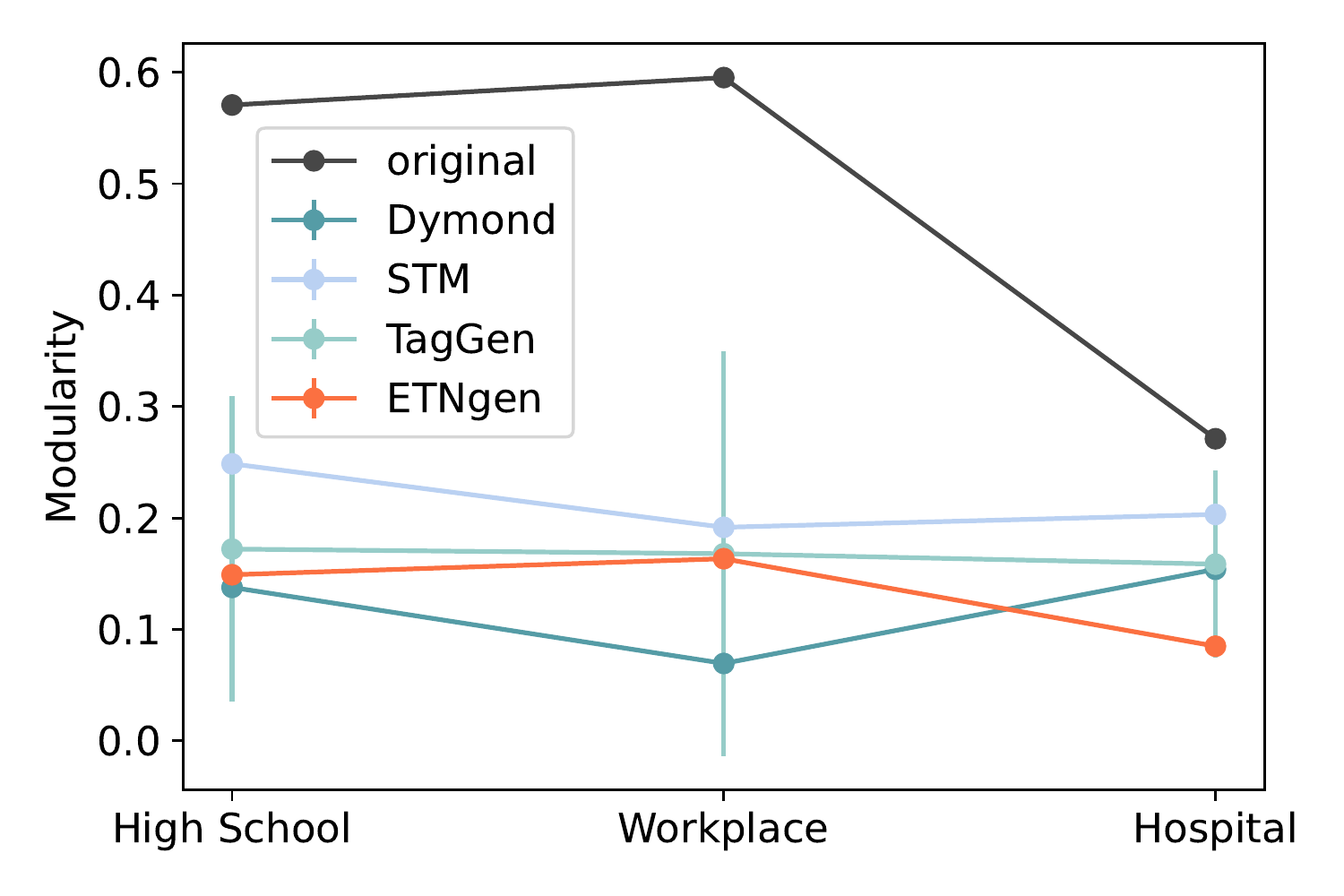}
    \caption{\changed{Detection of communities in the original and surrogate networks.}}
    \label{fig:si:com_det}
\end{figure}

\changed{As observed in the main text, \etngen is based on an egocentric perspective that does not allow to preserve some of the large- and meso-scale characteristics of spatial organization of the networks, like the existence of communities. In Figure~\ref{fig:si:com_det} we report a community detection analysis performed on the original and generated networks. We first obtained a partition of the aggregated networks using the Louvain algorithm~\cite{blondel2008fast} and then we computed modularity according to Clauset et al.~\cite{clauset2004finding}. We observe that, while the original networks are characterized by a certain level of modularity, none of the methods for surrogate networks is able to reproduce this feature.
}

\subsection{\changed{Motifs}}
\label{sec:si:motifs}
\begin{figure}[!h]
     \centering
     \includegraphics[width=\textwidth]{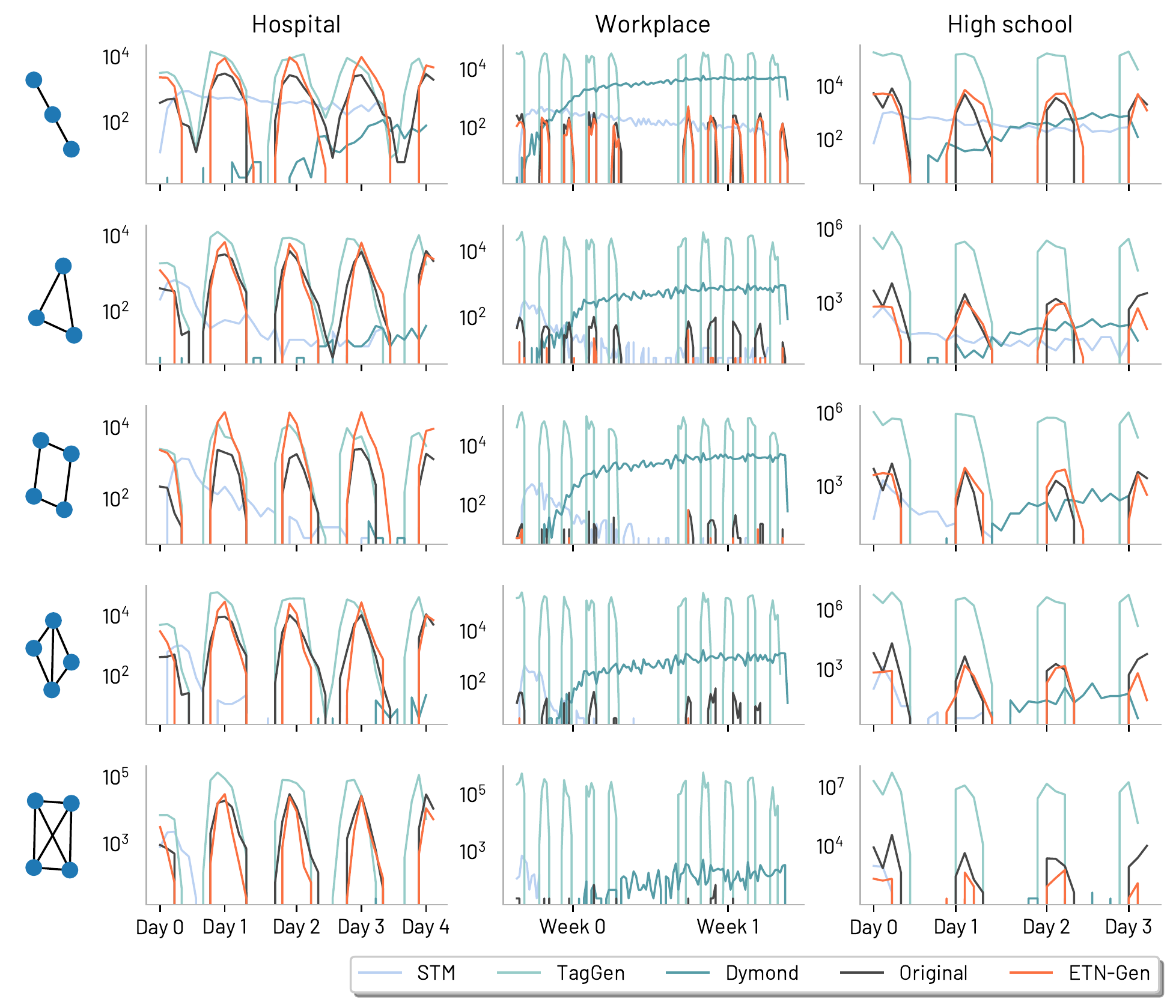}
    \caption{\changed{\textbf{Static motifs in time.} Appearance in time in the original and generated networks of the five static motifs represented on the left.}}
    \label{fig:si:static_motifs}
\end{figure}

\changed{
We tested the generated networks for the emergence of static and temporal motifs. We first investigated the presence of simple static motifs as those reported in Figure~\ref{fig:si:static_motifs} on the left.
We observe that \etngen allows the formation of these motifs with a similar amount to those appearing in the original graphs.}

\changed{
We also tested a posteriori the existence of egocentric temporal neighborhoods in the generated networks. We report in Table~\ref{tab:si:etn_motifs} the cosine distance between the number of occurrences of the egocentric temporal neighborhoods in original and generated networks. We limited the analysis to the most significant neighborhoods in the original graph. We used the concept of significant structures as reported in \cite{longa2021efficient} and inspired by \cite{milo2002network}.
As expected, the networks generated with \etngen accurately preserve these structures.}

\begin{table}[h!]
    \centering
    \begin{tabular}{l|c|c|c}
    & Hospital & Workplace & High school\\
    \hline
    \etngen     & 0.018	{\scriptsize($\pm$0.004)}& 	0.026{\scriptsize($\pm$0.007)}&	0.028	{\scriptsize($\pm$0.004)}\\
    \stm     & 0.463	{\scriptsize($\pm$0.092)}&	0.687{\scriptsize($\pm$0.048)}&	0.599	{\scriptsize($\pm$0.204)}\\
    \taggen     & 0.202	{\scriptsize($\pm$0.216)}&	0.487{\scriptsize($\pm$0.266)}&	0.626	{\scriptsize($\pm$0.312)}\\
    \dymond     & 0.734	{\scriptsize($\pm$0.145)}&	0.777{\scriptsize($\pm$0.001)}&	0.915	{\scriptsize($\pm$0.043)}
    \end{tabular}
    \caption{\changed{Temporal motifs: Cosine distance between original and generated networks in terms of egocentric temporal neighborhoods.}}
    \label{tab:si:etn_motifs}
\end{table}


\clearpage
\section{\changed{Different distances between distributions}}\label{sec:si:otherDist}

\changed{
In the Main Text we assessed the similarity between metrics distributions using the Kolmogorov-Smirnov distance. Here we report the definitions and results with three additional alternative distances: Kullback–Leibler divergence, Jensen–Shannon divergence, Earth mover's distance.}

\changed{
The Kullback-Leibler divergence\cite{Kullback51klDivergence}, given a population $X = {x_1,\dots,x_n}$ and two probability distributions $p(x)$ and $q(x)$, measures the information lost when $q(x)$ is used to approximate $p(x)$.
Formally, it is defined as:}
\begin{equation}
    D_{KL}(p(x)||q(x)) = \sum_{x \in X} p(x) \cdot \ln \dfrac{p(x)}{q(x)}
\end{equation}

\changed{While the Kullback-Leibler divergence is an asymmetrical measure, the Jensen-Shannon divergence\cite{menendez1997} provides a similar measure that is symmetric in nature. Formally it is defined as:}
\begin{equation}
    D_{JS}(p(x)||q(x)) = \dfrac{1}{2} D_{KL}(p(x)||M) +\dfrac{1}{2} D_{KL}(q(x)||M)
\end{equation}
\changed{Where $M$ is the mixture distribution of $p(x)$ and $q(x)$.}

\changed{The Earth mover's distance\cite{Ramdas2015OnWT}, also referred to as Wasserstein distance, quantifies the minimum amount of mass required to be moved to transform one distribution into another, effectively measuring their dissimilarity. Formally, it is defined as follow:}
\begin{equation}
    EMD(p(x),q(x)) = \min_{\lambda \in \tau(p(x),q(x))} \sum_{(x,y)\in X\times X} \lambda(x,y) \cdot c(x,y)
\end{equation}
\changed{
Where $\tau(p(x),q(x))$ represents the set of all possible joint distributions $\lambda$ that have $p(x)$ and $q(x)$ as marginals. $\lambda(x,y)$ is the amount of mass transported from point $x$ in $p(x)$ to point $y$ in $q(x)$, and $c(x,y)$ is the cost of transporting one unit of mass from $x$ to $y$.}
\\
\changed{In Tables \ref{tab:si:altre_dist_lh}, \ref{tab:si:altre_dist_in} and \ref{tab:si:altre_dist_hi}, we report these three distances together with the Kolmogorov-Smirnov distance for hospital, workplace and high school networks respectively. Despite variations in the magnitude of the distance or divergence observed among different measures, the relative rankings of the measures remain stable and the \etngen networks tendentially result the most close to the original ones.}

\begin{table}[h!]
\scriptsize
\begin{tabular}{l|cccc|cccc|cccc|cccc}
            & \multicolumn{4}{c|}{\textbf{Kolmogorov–Smirnov}} & \multicolumn{4}{c|}{\textbf{Jensen–Shannon}} & \multicolumn{4}{c|}{\textbf{Kullback–Leibler}} & \multicolumn{4}{c}{\textbf{Earth mover's distance}} \\ \hline
& E & S & T & D & E & S & T & D & E & S & T & D & E & S & T & D \\
Density     & \textbf{0.13}   & 0.67  & 0.19           & 0.55  & \textbf{0.05}  & 0.27 & 0.06          & 0.28 & \textbf{0.16}  & 10.94 & 0.81          & 11.15 & \textbf{0.02}   & 0.03    & 0.06           & 0.04   \\
Int. ind.   & \textbf{0.13}   & 0.43  & 0.19           & 0.57  & \textbf{0.05}  & 0.43 & 0.06          & 0.24 & \textbf{0.16}  & 7.76  & 0.81          & 8.14  & \textbf{4.95}   & 6.79    & 16.51          & 8.23   \\
New conv.   & \textbf{0.13}   & 0.43  & 0.31           & 0.56  & \textbf{0.03}  & 0.30 & 0.11          & 0.24 & \textbf{0.16}  & 5.95  & 2.02          & 7.92  & \textbf{2.22}   & 4.55    & 16.91          & 5.50   \\
Dur.        & 0.30            & 0.57  & \textbf{0.18}  & 0.56  & 0.07           & 0.20 & \textbf{0.02} & 0.20 & 0.27           & 4.45  & \textbf{0.21} & 7.89  & 0.70            & 0.57    & \textbf{0.22}  & 0.56   \\
GCC         & 0.40            & 0.41  & \textbf{0.20}  & 0.45  & 0.19           & 0.15 & \textbf{0.09} & 0.19 & 3.66           & 4.83  & \textbf{1.54} & 6.75  & 0.18            & 0.19    & \textbf{0.08}  & 0.21   \\
Ass.        & \textbf{0.25}   & 0.52  & 0.27           & 0.64  & \textbf{0.09}  & 0.34 & 0.11          & 0.40 & \textbf{1.49}  & 8.20  & 2.67          & 13.52 & \textbf{0.16}   & 0.32    & 0.20           & 0.61   \\
Con. com.   & \textbf{0.16}   & 1.00  & 0.22           & 1.00  & \textbf{0.03}  & 0.29 & 0.06          & 0.29 & \textbf{0.12}  & 1.20  & 1.29          & 1.20  & \textbf{3.28}   & 110.4   & 6.65           & 50.68  \\
H. clos.    & 0.50            & 0.71  & \textbf{0.29}  & 0.49  & 0.22           & 0.56 & \textbf{0.12} & 0.29 & 5.56           & 13.54 & \textbf{2.45} & 7.82  & 0.16            & 0.34    & \textbf{0.06}  & 0.20   \\
H betw.     & 0.25            & 0.43  & \textbf{0.23}  & 0.44  & \textbf{0.12}  & 0.19 & 0.18          & 0.24 & 2.65           & 4.18  & \textbf{2.16} & 6.53  & 0.03            & 0.03    & \textbf{0.02}  & 0.03   \\
W. h. betw. & 0.27            & 0.48  & \textbf{0.26}  & 0.51  & \textbf{0.13}  & 0.23 & 0.19          & 0.27 & 3.14           & 5.84  & \textbf{3.08} & 7.86  & 0.03            & 0.05    & \textbf{0.02}  & 0.04   \\
H. modu.    & 0.24            & 0.81  & \textbf{0.22}  & 0.32  & \textbf{0.17}  & 0.40 & 0.22          & 0.25 & \textbf{3.55}  & 11.76 & 5.26          & 8.69  & 0.09            & 0.47    & \textbf{0.06}  & 0.10   \\
H. S-met.   & \textbf{0.14}   & 0.44  & 0.29           & 0.65  & \textbf{0.10}  & 0.17 & 0.15          & 0.21 & \textbf{2.76}  & 7.41  & 3.67          & 9.02  & \textbf{1800}   & 3707    & 81590          & 3898   \\
H. aspl     & \textbf{0.32}   & 0.67  & 0.33           & 0.51  & 0.18           & 0.31 & \textbf{0.15} & 0.25 & 3.62           & 7.60  & \textbf{2.33} & 6.96  & 0.34            & 1.42    & \textbf{0.28}  & 0.73  
\end{tabular}
\caption{\changed{Topological similarity in the Hospital dataset comparing alternative distance measures: Kolmogorov-Smirnov, Jensen–Shannon, Kullback–Leibler, and Earth mover's distance. E, S, T, and D represent \etngen, STM, \taggen, and \dymond, respectively.}}
\label{tab:si:altre_dist_lh}
\end{table}

\begin{table}[h!]
\scriptsize
\begin{tabular}{l|cccc|cccc|cccc|cccc}
            & \multicolumn{4}{c|}{\textbf{Kolmogorov–Smirnov}} & \multicolumn{4}{c|}{\textbf{Jensen–Shannon}} & \multicolumn{4}{c|}{\textbf{Kullback–Leibler}} & \multicolumn{4}{c}{\textbf{Earth mover's distance}} \\ \hline
& E & S & T & D & E & S & T & D & E & S & T & D & E & S & T & D \\
Density     & \textbf{0.05} & 0.68          & 0.20 & 0.68 & \textbf{0.01} & 0.26          & 0.08 & 0.33 & \textbf{0.09} & 5.96          & 1.46  & 1.56          & \textbf{0.01} & 0.05  & 0.02          & 0.02 \\
Int. ind.   & \textbf{0.05} & 0.68          & 0.20 & 0.67 & \textbf{0.01} & 0.43          & 0.08 & 0.32 & \textbf{0.09} & 15.58         & 1.46  & 1.45          & \textbf{0.30} & 1.55  & 6.54          & 6.87 \\
New conv.   & \textbf{0.11} & 0.27          & 0.64 & 0.66 & \textbf{0.02} & 0.14          & 0.31 & 0.28 & \textbf{0.45} & 1.05          & 5.06  & 1.18          & \textbf{0.49} & 0.93  & 18.0          & 5.95 \\
Dur.        & \textbf{0.10} & 0.35          & 0.19 & 0.34 & \textbf{0.03} & 0.12          & 0.04 & 0.12 & \textbf{0.40} & 2.52          & 0.43  & 3.72          & \textbf{0.11} & 0.298 & 0.19          & 0.30 \\
GCC         & 0.03          & \textbf{0.01} & 0.13 & 0.15 & 0.02          & \textbf{0.01} & 0.06 & 0.04 & 0.23          & \textbf{0.13} & 0.40  & 0.14          & 0.03          & 0.005 & \textbf{0.02} & 0.07 \\
Ass.        & \textbf{0.22} & 0.44          & 0.50 & 0.29 & \textbf{0.08} & 0.29          & 0.29 & 0.12 & 1.49          & 4.88          & 4.73  & \textbf{0.54} & \textbf{0.24} & 0.347 & 0.36          & 0.19 \\
Con. com.   & \textbf{0.05} & 1.00          & 0.20 & 0.69 & \textbf{0.01} & 0.50          & 0.07 & 0.35 & \textbf{0.05} & 2.66          & 1.46  & 5.56          & \textbf{0.23} & 107   & 4.66          & 7.01 \\
H. clos.    & \textbf{0.07} & 0.61          & 0.31 & 0.83 & \textbf{0.02} & 0.46          & 0.21 & 0.51 & \textbf{0.16} & 14.13         & 5.45  & 4.89          & \textbf{0.01} & 0.08  & 0.10          & 0.21 \\
H betw.     & \textbf{0.04} & 0.62          & 0.25 & 0.84 & \textbf{0.02} & 0.06          & 0.07 & 0.45 & \textbf{0.17} & 0.30          & 0.31  & 2.22          & 0.02          & 0.01  & \textbf{0.01} & 0.03 \\
W. h. betw. & \textbf{0.04} & 0.63          & 0.26 & 0.84 & \textbf{0.02} & 0.07          & 0.08 & 0.44 & \textbf{0.18} & 0.36          & 0.38  & 2.18          & 0.02          & 0.02  & \textbf{0.01} & 0.03 \\
H. modu.    & \textbf{0.18} & 0.56          & 0.77 & 0.71 & \textbf{0.07} & 0.15          & 0.50 & 0.37 & 1.37          & \textbf{1.09} & 13.87 & 3.64          & \textbf{0.03} & 0.11  & 0.32          & 0.17 \\
H. S-met.   & \textbf{0.08} & 0.70          & 0.30 & 0.80 & \textbf{0.03} & 0.36          & 0.09 & 0.20 & \textbf{1.00} & 3.11          & 3.22  & 1.58          & \textbf{21}   & 63    & 20484         & 2212 \\
H. aspl     & \textbf{0.04} & 0.63          & 0.20 & 0.82 & \textbf{0.05} & 0.35          & 0.11 & 0.48 & \textbf{0.84} & 13.60         & 2.96  & 5.58          & \textbf{0.07} & 1.56  & 0.28          & 2.55
\end{tabular}
\caption{\changed{Topological similarity in the Workplace dataset comparing alternative distance measures: Kolmogorov-Smirnov, Jensen–Shannon, Kullback–Leibler, and Earth mover's distance. E, S, T, and D represent \etngen, STM, \taggen, and \dymond, respectively. }}
\label{tab:si:altre_dist_in}
\end{table}

\begin{table}[h!]
\scriptsize
\begin{tabular}{l|cccc|cccc|cccc|cccc}
            & \multicolumn{4}{c|}{\textbf{Kolmogorov–Smirnov}} & \multicolumn{4}{c|}{\textbf{Jensen–Shannon}} & \multicolumn{4}{c|}{\textbf{Kullback–Leibler}} & \multicolumn{4}{c}{\textbf{Earth mover's distance}} \\ \hline
& E & S & T & D & E & S & T & D & E & S & T & D & E & S & T & D \\
Density     & \textbf{0.09} & 0.55          & 0.28          & 0.30          & \textbf{0.03} & 0.18          & 0.10 & 0.16 & \textbf{0.56} & 8.03          & 0.84          & 3.26          & \textbf{0.04} & 0.11          & 0.50          & 0.11          \\
Int. ind.   & \textbf{0.09} & 0.56          & 0.28          & 0.27          & \textbf{0.03} & 0.19          & 0.10 & 0.08 & \textbf{0.56} & 3.33          & 0.84          & 1.66          & \textbf{3.41} & 7.85          & 39.65         & 6.18          \\
New conv.   & \textbf{0.16} & 0.27          & 0.66          & 0.18          & \textbf{0.07} & 0.09          & 0.31 & 0.08 & \textbf{1.67} & 3.35          & 4.87          & 2.55          & \textbf{5.56} & 8.31          & 74.41         & 6.98          \\
Dur.        & 0.24          & 0.38          & \textbf{0.20} & 0.37          & \textbf{0.06} & 0.14          & 0.04 & 0.14 & 0.34          & 2.94          & \textbf{0.28} & 4.02          & 0.54          & 0.39          & \textbf{0.24} & 0.39          \\
GCC         & 0.14          & \textbf{0.08} & 0.18          & 0.13          & 0.07          & \textbf{0.05} & 0.13 & 0.05 & 1.59          & \textbf{1.25} & 1.90          & 0.78          & 0.06          & \textbf{0.05} & 0.06          & 0.09          \\
Ass.        & 0.33          & 0.70          & 0.54          & \textbf{0.32} & \textbf{0.13} & 0.45          & 0.26 & 0.18 & 2.25          & 8.70          & 5.98          & \textbf{1.43} & 0.25          & 0.41          & 0.36          & \textbf{0.24} \\
Con. com.   & \textbf{0.10} & 1.00          & 0.28          & 0.99          & \textbf{0.04} & 0.34          & 0.10 & 0.61 & \textbf{0.53} & 1.93          & 1.24          & 16.83         & \textbf{2.29} & 247           & 15            & 63            \\
H. clos.    & \textbf{0.22} & 0.49          & 0.40          & 0.58          & \textbf{0.14} & 0.43          & 0.27 & 0.30 & \textbf{3.78} & 14.61         & 7.71          & 4.50          & \textbf{0.05} & 0.06          & 0.12          & 0.20          \\
H betw.     & 0.21          & 0.50          & \textbf{0.20} & 0.55          & \textbf{0.10} & 0.13          & 0.16 & 0.29 & \textbf{1.47} & 2.99          & 3.73          & 2.25          & 0.08          & 0.07          & \textbf{0.05} & 0.38          \\
W. h. betw. & 0.21          & 0.50          & \textbf{0.19} & 0.52          & \textbf{0.09} & 0.14          & 0.16 & 0.27 & \textbf{1.67} & 3.49          & 3.72          & 2.13          & 0.09          & 0.07          & \textbf{0.06} & 0.37          \\
H. modu.    & \textbf{0.29} & 0.46          & 0.73          & 0.46          & 0.31          & \textbf{0.27} & 0.52 & 0.35 & 7.95          & \textbf{6.94} & 15.21         & 8.61          & \textbf{0.09} & 0.14          & 0.34          & 0.18          \\
H. S-met.   & \textbf{0.12} & 0.56          & 0.35          & 0.42          & \textbf{0.07} & 0.08          & 0.10 & 0.09 & \textbf{1.84} & 3.67          & 2.92          & 2.65          & \textbf{260}  & 406           & 5966          & 357           \\
H. aspl     & \textbf{0.18} & 0.56          & 0.23          & 0.44          & \textbf{0.14} & 0.42          & 0.22 & 0.29 & \textbf{3.44} & 13.68         & 6.05          & 3.68          & \textbf{0.43} & 1.66          & 0.44          & 1.15         
\end{tabular}

\caption{\changed{Topological similarity in the High school dataset comparing alternative distance measures: Kolmogorov-Smirnov, Jensen–Shannon, Kullback–Leibler, and Earth mover's distance.  E, S, T, and D represent \etngen, STM, \taggen, and \dymond, respectively.}}
\label{tab:si:altre_dist_hi}
\end{table}



\end{document}